1
2
3

4 **Earthquake Depth-Energy Release:  Thermomechanical**

5 **Implications for Dynamic Plate Theory**

6
7
8
9
10
11
12
13

14 Running title: EARTHQUAKE DEPTH-ENERGY RELEASE

15 Words abstract: 290

16 Words text: 13552

17 References: 93

18 Tables: 6

19 Figures: 9

20
21
22
23
24
25
26
27
28

29 R. L. Patton

30 School of Environmental Sciences

31 Washington State University

32 Pullman, WA  99164

33 rpatton@wsu.edu

35 **Abstract.** Analysis of the global centroid-moment tensor catalog reveals significant regional variations of

36 seismic energy release to 290 km depth. The largest variations, with direction from the baseline indicated

37 using plus and minus signs, and in decreasing order, occur at 14-25 km depths in continental transform (+),

38 oceanic ridge/transform (+), continental rift (+), Himalayan-type (+), island arc-type (-) and Andean-type (-

39 ) margins. At 25-37 km depths, variations one-fifth the size occur in continental rift (+), island arc-type

40 (+), Andean-type, (-), Himalayan-type (-), oceanic ridge/transform (-), and continental transform (-)

41 margins. Below 37-km depth, variations one-tenth the size occur in Andean-type and Himalayan-type

42 margins to depths of about 260 km. Energy release in island arc-type margins closely tracks the baseline to

43 the maximum depth of earthquakes at 699 km. The maximum depth of earthquakes in Andean-type and

44 Himalayan-type margins is 656 and 492 km, respectively, while in divergent and transform margins it is

45 about 50 km. These variations reflect radial and lateral contrasts in thermomechanical competence,

46 consistent with a shear-dominated non-adiabatic boundary layer some 700-km thick, capped by denser

47 oceanic lithosphere as much as 100 km thick, or lighter continental tectosphere 170 to 260 km thick. Thus,

48 isobaric shearing at fractally-distributed depths likely facilitates toroidal plate rotations while minimizing

49 global energy dissipation. Shear localization in the shallow crust occurs as dislocations at finite angles with

50 respect to the shortening direction, with a 30 degree angle being the most likely. Consequently, relatively

51 low-angle (~30º) reverse faults, steep (~60º) normal faults, and triple junctions with orthogonal or

52 hexagonal symmetry are likely to form in regions of crustal shortening, extension, and transverse motion,

53 respectively. Thermomechanical theory also predicts adiabatic conditions in the mantle below about 1000-

54 km depth, consistent with observed variations in bulk sound speed.

**1.0 Introduction**

A century ago, the solid mechanical properties of rocks factored highly in the development of the isostatic hypothesis, which holds that Earth's outer layers, the crust and lithosphere, are strong and capable of supporting loads for geological time periods (Watts, 2001). Vertical motions of the lithosphere, in response to progressive sedimentary loading, volcano construction, and the advance of ice sheets, are accommodated by deformation of a subjacent weak asthenosphere (Barrell, 1914). Subsequent removal of loads, due to erosion or ablation, results in the rebound of formerly depressed areas. Thus a dynamic equilibrium between Earth's shape, topography, and external gravity field is maintained.

In principle, the relative horizontal motions of continents suggested by Wegener (1966) are consistent with the presence of the asthenosphere. Still, the mechanism he proposed, continental 'rafts' plowing undeformed through oceanic crustal 'seas' under the influence of centrifugal forces, is at odds with the known contrast between stronger mafic oceanic crustal rocks and weaker felsic continental ones. Consequently, wide acceptance of continental drift had to await the development of the sea floor spreading hypothesis (Dietz, 1961; Hess, 1962), and recognition of Benioff (1954) zones as down going extensions of oceanic lithosphere. Thus, the continents appear to be embedded within blocks of mobile lithosphere, the horizontal motion of which is decoupled from the deeper mantle by the asthenosphere, and accommodated by its progressive creation and destruction at oceanic ridges and trenches, respectively (Isacks et al., 1968). Lithosphere is neither created nor destroyed at transform faults, where the relative motion of adjacent blocks is parallel to the fault (Wilson, 1965).

In the late 1960's, the kinematic theory of plate tectonics (Le Pichon, 1968; McKenzie and Morgan, 1969; McKenzie and Parker, 1967; Morgan, 1968; Wilson, 1965) successfully explained global map patterns of earthquake foci and first motions (Isacks et al., 1968), and systematic variations of sea floor age across active oceanic ridges (Vine and Matthews, 1963), via the relative horizontal motion of rigid spherical caps. Attempts to reconcile this new theory with observed patterns in continental geology followed soon after (Atwater, 1970; Dewey and Bird, 1970), but the rigidity assumption has never sat well with the overall width of, and diffuse nature of deformation in, mountain belts. Other limitations of plate theory include its 83    inability to account for known differences in the density and strength of crustal rocks between oceans and

84    continents, and the various thickness estimates for the crust (Rudnick and Gao, 2003) and oceanic

85    lithosphere (Parsons and Sclater, 1977), as well as stable cratons and shields (Artemieva, 2009; Artemieva

86    et al., 2004; Artemieva and Mooney, 2001). Moreover, it does not predict the existence of seismicity, but

87    rather uses it to define the boundaries of, and relative motions between, its plate-like blocks. These

88    boundaries apparently act as long-lived zones of weakness in the global plate system. The dynamic basis

89    for plate theory and its relation to motions in the deeper mantle remain poorly understood (Bercovici et al.,

90    2000; McKenzie, 1969; Tackley, 2000).

91

92    In order to simulate horizontal (toroidal) rotations of plate-like blocks many variations of the standard Earth

93    model (Bercovici et al., 2000) incorporate *ad hoc* mechanisms for shear localization. However, because

94    these mechanisms are based on temperature-dependent variations in viscosity, and ideal viscous materials

95    do not manifest shear bands or dislocations, they cannot account for seismicity. This also holds for power

96    law viscous, i.e. Reiner-Rivlin, materials. Although these materials can produce shear bands for exponents

97    $2 \leq n < \infty$, they cannot simultaneously propagate shear waves, and hence are rheological fluids (Patton and

98    Watkinson, 2010). Therefore, a major stumbling block to further progress on the dynamic plate problem is

99    not a lack of data, but rather a deficiency in applied rheological theory.

100

101   Since the mid-1960's, space-based geodesy has provided global gravity and topography datasets with ever

102   increasing quality and resolution. Today, the tracking of satellite orbits constrains the low-frequency

103   harmonics of Earth's external gravity field, with half-wavelengths ($\lambda_{\frac{1}{2}}$) greater than about 200 km

104   (spherical harmonic degrees *l < 100)*. Analysis of these data, combined with gravimetric survey and global

105   topography data (Amante and Eakins, 2009), show that Earth's gravity and topography are correlated (*r =

106   0.6-0.7)* for $\lambda_{\frac{1}{2}} < 1000$ *(l > 20)* (Wieczorek, 2007), consistent with isostasy at a regional level (Watts,

107   2001). It is therefore common to hear tectonic specialists talking about the effective elastic thickness of the

108   crust and lithosphere. The gravity-topography correlation diminishes rapidly for $l \leq 20$, which implies that

109   vertical (poloidal) motions in the mantle actively support the longest wavelength gravity and geoid

110   anomalies (McKenzie, 1967).

111

112 An outstanding question for dynamic Earth models, and the standard model in particular, is whether long-

113 wavelength gravity anomalies can be associated unequivocally with sub-lithospheric mantle convection

114 (Kaula, 1972; Steinberger et al., 2010) and so-called dynamic topography (Hager, 1984; Panasyuk and

115 Hager, 2000). Here I show that key features of Earth's gravity-topography correlation and admittance can

116 be explained using a thermomechanical boundary layer hypothesis, which constrains the possible range of

117 wavelengths associated with standard convection to $\lambda_{\frac{1}{2}} > 2500$ km ($l < 8$). This limit is far larger than the

118 $\lambda_{\frac{1}{2}} > 650$ km ($l < 31$) one identified by Steinberger *et al* (2010) based on empirical 'downward

119 continuation' of gravity spectra and crustal thickness modeling, and wholly consistent with the scale of

120 robust heterogeneities revealed by seismic tomography (Dziewonski et al., 1977; Dziewonski and

121 Woodward, 1992; Gu et al., 2001).

122

123 Establishment of the Global Positioning System (GPS) in the early 1990's makes it possible to account for

124 ever longer-period neotectonic deformations. These new datasets have been used recently to estimate

125 density-normalized rates of mechanical energy dissipation $\chi$ [m$^2$s$^{-1}$] in plate boundary zones, which fall in

126 the range $4.2 \leq -log\ \chi \leq 5.2$, and also in intraplate regions, which fall in the range $5.3 \leq -log\ \chi \leq 7.0$ (Patton

127 and Watkinson, 2010). As shown later in the paper, similar estimates of $\chi$ for spontaneous-failures in load

128 hold experiments on the Mt. Scott granite (Katz and Reches, 2002, 2004) fall in the range $2.3 \leq -log\ \chi \leq$

129 $5.6$. The overlap in these estimates suggests that joint analysis of neotectonic survey and rock mechanics

130 data could constrain natural rates of energy dissipation in the broad deforming plate boundary regions,

131 which have been so problematic for the kinematic theory.

132

133 Furthermore, by comparing these mechanical dissipation estimates to measured rates of thermal energy

134 dissipation $\kappa$ [m$^2$s$^{-1}$] in silicates, e.g. $5.9 \leq -log\ \kappa \leq 6.3$ (Clauser and Huenges, 1995; Vosteen and

135 Schellschmidt, 2003), it is possible to define the thermomechanical competence of solid earth materials by

136 the ratio $\kappa/\chi$. Note, for values of $\kappa/\chi < 1$, the rate of mechanical dissipation is higher than thermal

137 dissipation, so that as a material deforms it carries its heat along. Conversely, values of $\kappa/\chi > 1$ imply that

138 the rate of thermal dissipation is higher than mechanical dissipation, so that heat will be conducted readily

139 through a material while it deforms only slightly. This, in a nutshell, is the thermomechanical rigidity

140 hypothesis (Patton and Watkinson, 2010), which is used here to interpret observed variations in seismic

141 energy release. This definition of rigidity retains its practical meaning as a resistance to shape changes,

142 while shedding the highly idealized kinematic connotation of no internal deformation under loading.

143

144 Mechanical properties also factor highly in seismological models for wave propagation, and are critical for

145 inferring first-order planetary structure, like the existence and nature of the mantle and core. As is well-

146 known, the lack of shear wave propagation through the outer core is consistent with its existence in the

147 liquid phase, while the propagation of shear waves in the mantle and inner core is consistent with their

148 existence in the solid phase. Additionally, the attenuation of surface waves in the shallow asthenosphere,

149 coincident with the low velocity zone, is consistent with the presence of a partial melt phase (Presnall and

150 Gudfinnsson, 2008), although solid-state anelasticity also plays a role (Jackson et al., 2005; Karato, 1993).

151

152 Defining thermomechanical competence by the ratio $\kappa/\chi$ has the advantage of being testable, and free of the

153 confounding aspects of the effective elastic thickness estimated in regional isostasy studies (Watts, 2001).

154 For example, if the entire mantle propagates shear waves and is therefore an elastic solid, how can it make

155 sense to further define plate-like blocks by their effective elastic thickness? Granted, the seismologic and

156 isostatic models serve different purposes, and hence their use of the term 'elastic' need not be consistent.

157 Nevertheless, a consistent definition of material competence is required for any dynamic plate theory, and I

158 shall use thermomechanical competence throughout.

159

160 Adams and Williamson (1923), and later Birch (1952), used the relation

161
$$K_s/\rho = V_p^2 - (4/3)V_s^2 = V_\phi^2 \qquad (1)$$

162 where $K_s$ is adiabatic bulk incompressibility and $\rho$ is density, to study the thermodynamics of the mantle

163 based on observed radial variations in compressional $V_p$, shear $V_s$, and bulk sound $V_\phi$ wave speeds.

164 Birch inferred that the mantle below about 900-km depth was chemically homogeneous and adiabatic,

165 while between 200- and 900-km depths it was not. Furthermore, he was among the first to suggest the 166    importance of mineral phase changes in the upper transitional zone.  Seismological data collected since the

167    1950's have done nothing but reinforce Birch's insights (Dziewonski and Anderson, 1981).  Clearly, there

168    is something unusual about the upper mantle that must be understood thermodynamically.

169

170    Proponents of the standard Earth model often point to the apparent adiabaticity of the lower mantle as

171    evidence for its vigorous convection.  Although an adiabatic gradient can be maintained by the efficient

172    mixing of material in viscous convection cells at high Rayleigh number (Turcotte and Oxburgh, 1967),

173    adiabaticity also prevails in the lower reaches of a solid thermomechanical mantle, whether it is convecting

174    or not.  Furthermore, because the thermal expansivity of such mantle is inversely proportional to

175    temperature, any convection process in the lower, hotter, mantle would likely be rather sluggish compared

176    to that in the upper mantle.  These conclusions, based on the statistical thermodynamics of non-linear

177    elastic solids, are consistent with those based on the compressibility and rigidity of likely lower mantle

178    mineral species (Anderson, 1989, 2007).

179

180    In summary, any dynamic plate theory should predict the plate-like nature of Earth's outer shell, provide

181    self-consistent mechanisms for its toroidal motion and spontaneous localization of shear, and explain the

182    observed spectral correlation and admittance of Earth's topography and gravity.  In doing so, the nature of

183    coupling between the plate-like blocks at the surface and poloidal motions in the mantle should be made

184    clear.  Proponents of the standard Earth model emphasize the insignificance of crustal and lithospheric

185    strength on the length and time scales of geodynamics, and point to estimates of mantle viscosity based on

186    glacial-isostatic rebound as evidence in favor of their view.  This stance is necessary, only because the

187    rheological theory upon which the standard model is based does not account for material strength or shear

188    localization.

189

190    Here, I adopt the complimentary stance, and ask what are the length and time scales over which the force of

191    gravity acts within the body of a solid thermomechanical planet.  In deriving the dynamic rescaling theorem

192    for deforming differential grade-2 (DG-2) solids (Patton and Watkinson, 2010), I have come to realize that

193    rock strength always matters, not only for plate theory but also for the entire mantle.  Thus, the 194  confounding semantics of rock 'viscosity' are moot, because a clear theoretical distinction exists between

195  the deformation of thermomechanical solids on the one hand, and thermoviscous fluids on the other.

196  Consequently, the density-normalized rates of mechanical energy dissipation measured in rock mechanics

197  experiments can be applied immediately to the dynamic plate problem, regardless of whether they are

198  termed kinematic viscosities or mechanical diffusivities. Henceforth, I shall adopt the latter term, in an

199  effort to discourage further semantic confusion.

200

201  This study interprets earthquake depth-energy release patterns for tectonically regionalized data from the

202  global centroid-moment tensor (CMT) catalog (Ekstrom and Nettles, 2011), using insights from the

203  thermomechanical theory of non-linear elastic DG-2 materials (Patton, 1997; Patton et al., 2000; Patton and

204  Watkinson, 2005; Patton and Watkinson, 2010, in review). These materials exhibit both distributed

205  harmonic and localized shear band modes of deformation as a function of $\kappa/\chi$. Harmonic modes alone are

206  possible for $0 < \kappa/\chi < \frac{1}{2}$, while both harmonic and shear band modes are possible for $\kappa/\chi > \frac{1}{2}$. This

207  transition occurs on a domain of thermomechanical competence lower than that associated with rigidity.

208  Consequently, theory predicts that even rigid materials can suffer irreversible shear band deformation.

209  Because shear localization liberates strain energy and increases entropy, it is reasonably identified with

210  earthquake faulting and damage. The spontaneous nature of this transition therefore offers a unique

211  opportunity to incorporate global seismicity into dynamic plate theory in a self-consistent manner.

212

213  Section 2.0 of the paper examines the map, magnitude, and depth distribution of seismicity as sampled by

214  the CMT catalog, and explains the method and rationale for computing earthquake depth-energy release

215  curves. It compares and contrasts depth variations in this signal using a six-fold classification of tectonic

216  margins. This section provides some background on the CMT inversion process, as compared to typical

217  earthquake location methods, so that the relevance of CMTs to dynamic plate theory is clear.

218

219  Section 3.0 outlines a statistical thermodynamic theory for strained inhomogeneous elastic and self-

220  gravitating matter configurations (Lavenda, 1995), which place severe constraints on the slope and shape of

221  the energy density as a function of entropy and length. This theory makes specific predictions about

222  temperature and pressure variations and the temperature dependence of thermal expansivity in these

223  materials.  Alone, these findings are consistent with expected variations of pressure and temperature in

224  terrestrial planets, but offer no insight into observed variations of seismic energy release with depth in the

225  Earth.  Consequently, they provide essential foil for the thermomechanics of shear localization in DG-2

226  materials.

227

228  Section 4.0 outlines a statistical thermomechanical theory for DG-2 materials.  It builds upon the

229  consistency of the slope and shape of the distributed energy threshold ($\psi^D$) for DG-2 materials with the

230  constraints derived in Section 3, and its interpretation as an elastic strain-energy function (Patton and

231  Watkinson, 2005), to place earlier published work on these ideal materials in a very general and

232  geologically useful context.  Domains of $\kappa/\chi$ for which harmonic and shear band modes of deformation are

233  possible are deduced using incipient modes analysis.

234

235  Section 5.0 applies thermomechanical theory to the interpretation and correlation of published data from

236  rock mechanics experiments on the Mt. Scott Granite (2002, 2004).  Not only does the localization curve

237  ($\psi^L$) on the lower-competence harmonic domain neatly divide the sample population into macroscopically

238  failed and un-failed groups, but it also correlates with post-loading observations of microscopic crack and

239  macroscopic shear angles on the mixed harmonic-shear band domain at higher competence.  The dynamic

240  rescaling of lengths during sample failure is likely (Patton and Watkinson, 2010), given that the pre- and

241  post-failure observations span the non-convex portion of the localization curve (Hobbs et al., 2011).

242  Dynamic shear failure in these rock samples therefore minimizes energy dissipation in the combined

243  sample-load frame system.  Data from three samples suffering spontaneous shear localization are used to

244  estimate mechanical diffusivity $\chi$.

245

246  Section 6.0 expands upon the isobaric shearing hypothesis (Patton and Watkinson, 2010) and its utility for

247  the interpretation of Earth structure.  Predicted depths to these theoretical shears, empirically calibrated via

248  least-squares minimization of ThERM to PREM (Patton, 2001; Patton and Watkinson, 2009, 2010), are

249  consistent with observed variations of earthquake depth-energy release.  Furthermore, their fractal depth

250    distribution suggests that Earth's dynamic plate system globally minimizes energy dissipation. The

251    structure of the lithosphere and asthenosphere, together, comprise a thermomechanical boundary layer

252    adjacent of the surface of the planet.

253

254    Section 7.0 discusses the implications of these findings for the interpretation of rock mechanics data,

255    pressure-temperature-time data from structural and metamorphic studies in orogens, likely source depths

256    for common intrusive and extrusive rocks, the lateral variability of seismic wave speeds in the upper mantle

257    and crust, and the spectral correlation and admittance of Earth's gravity and topography.

258

259    Section 8.0 concludes the paper with the notion that coupled toroidal-poloidal motions of Earth's plate-like

260    blocks represent a top-down mode of convection peculiar to solid thermomechanical planets. Given the

261    generality of the theory upon which these conclusions rest, and its remarkable correlation with datasets

262    from a wide range of fields, it is likely that the absence of plate-like convection on other terrestrial planets

263    can be attributed to the small size of the sample population, as well as the lack of liquid water on the known

264    examples. Water and other volatile species tend to weaken, i.e. increase the mechanical diffusivity of,

265    common rocks and minerals.

266

267    **2.0 Earthquake depth-energy release**

268    <u>2.1 Locating earthquakes</u>

269    Since about 1970, earthquake hypocenters have been routinely and rapidly located using telemetric

270    monitoring networks established by the United States Geological Survey, and many other agencies, for the

271    purpose of mitigating earthquake hazards. These networks typically use body wave first motions to

272    triangulate event locations within radially-symmetric elastic Earth models (Kennett and Engdahl, 1991),

273    although for the past 15 years or so depth phases also have been used (Engdahl et al., 1998). Body waves

274    propagate freely through solids with lateral extents much greater than the wavelength of the waves

275    themselves. They are highly sensitive to variations in density and elastic moduli at the scale of about one-

276    half their wavelength. Consequently, the average error for hypocenter depth in standard catalogs, like the

277  Preliminary Determination of Epicenters, is about 14 km (Kennett and Engdahl, 1991).  Hypocenter depths

278  for 20646 earthquakes are shown in Figure 1a.

279

280  By the time of the 1960 May 22 $M_w$9.5 Chile earthquake, which started the planet ringing like a bell,

281  instrumentation had been developed that allowed the first precise studies of Earth's free oscillations.  These

282  so called normal modes have periods as long as 54 minutes, and their observation spawned a whole new

283  area of seismological research.  Subsequent theoretical developments showed that the precise location of

284  earthquakes, and their detailed dislocation characteristics, could be determined within the body of an Earth

285  model by summing these normal modes (Dziewonski et al., 1981; Dziewonski and Woodhouse, 1983;

286  Ekstrom et al., 2012).  This CMT method accounts for long-period energy that is not possible via first

287  motions of body waves alone, and does not require a radially-symmetric reference model.  Furthermore, it

288  is insensitive to lateral heterogeneity at scales less than half the wavelength of a given normal mode (Luh,

289  1975; Madariaga, 1972).

290

291  Since the early 1980's, recordings of well-observed earthquakes have been routinely inverted for their

292  CMT characteristics via normal modes summation (Dziewonski et al., 1981; Dziewonski and Woodhouse,

293  1983; Ekstrom et al., 2012).  In this process the triangulated hypocenter is taken as the initial estimated

294  location of a shear dislocation, which is then refined by accounting for the energy contained in the gravest

295  and progressively higher frequency modes.  Centroid depths for 20646 of these are shown in Figure 1b.

296  The global CMT catalog, begun by Adam Dziewonski and co-workers at Harvard, is now maintained under

297  the CMT project (Ekstrom and Nettles, 2011) at Lamont-Doherty Earth Observatory.

298

299  CMT inversion initially took into account long-period body waves (Dziewonski et al., 1981), with peak

300  spectral energy between about 16.7 to 20 mHz, and Earth's free oscillations with frequencies lower than 7.4

301  mHz (Dziewonski and Woodhouse, 1983), the so-called 'mantle waves'.  However, since 2004 normal

302  modes with frequencies in the intermediate 7.4 to 16.7 mHz range, which are strongly affected by lateral

303  Earth structure (Luh, 1975; Madariaga, 1972), also have been used in routine inversions (Ekstrom et al.,

304  2012).  As a result the practical lower limit on event magnitude, initially $M_w > 5.5$ and associated with a

305  frequency cut-off above about 22 mHz, has decreased to about $M_w > 5.0$.

306

307  Reference models used by the CMT project include the radially-symmetric PREM (Dziewonski and

308  Anderson, 1981) and shear attenuation model QL6 (Durek and Ekstrom, 1996), and a mantle heterogeneity

309  model called SH8U4L8 (Dziewonski and Woodward, 1992).  The latter model accounts for lateral

310  variations in shear wave speeds at half-wavelengths as small as about 2500 km ($l \leq 8$), in four layers of the

311  upper mantle, and eight layers of the lower mantle.  Lateral variations of this scale represent the most

312  robust deviations from spherical symmetry, as imaged by global tomographic models (Dziewonski et al.,

313  1977; Gu et al., 2001).  Furthermore, the majority of shear attenuation occurs in the upper mantle, above

314  about 670-km depth (Durek and Ekstrom, 1996; Romanowicz, 1994; Romanowicz, 1995; Widmer et al.,

315  1991).

316

317  2.2 Effects of CMT relocation

318  The CMT catalog, downloaded from www.globalcmt.org, includes data for 30872 earthquakes recorded

319  during the period January 1976 through December 2010.  Of these only 20646 were relocated by the CMT

320  procedure (Ekstrom and Nettles, 2011), and therefore have quantitative error estimates.  The remaining

321  events either had their focal depths fixed by an analyst, or constrained by an inversion of short-period data.

322  The consistent algorithmic treatment of the 20646 relocated events, and the fact that CMTs estimate the

323  total work done by a seismic dislocation, make them ideal for the tectonic analysis presented here.

324

325  Mean standard errors for CMT latitude, longitude, and depth are 0.042º, 0.047º, and 2.82 km, respectively.

326  The depth error estimate defines the minimum thickness of a filter used to smooth the earthquake depth-

327  energy release curves presented here.  A filter smaller than this tends to display more noise than a larger

328  one, while larger filters discard potentially interpretable depth signal in the dataset, at least at typical crustal

329  depths.  Depth-energy release curves shown here were smoothed using either 3-km or 10-km filters.

330

331    Initial hypocenters and relocated centroids are distributed differently with both magnitude (Table 1) and

332    depth (Table 2).  Centroids generally have higher moment magnitudes than their associated hypocenters,

333    consistent with the fact that longer-period energy is accounted for in their estimation.  All 20646 relocated

334    events appearing in the catalog are accounted for in the moment release studies presented here.  Most of the

335    seismic signal is present in the $M_w$=5-6 band, with significant signal also in the $M_w$=4-5 and $M_w$=6-7 bands.

336    The depth distributions of hypocenters and centroids are significantly different in the crust, above about 37-

337    km depth.  Small differences in map locations exist as well, but these do not significantly alter the global

338    map pattern of seismicity, and are not discussed further.

339

340    Figure 1 shows differences in the depth distribution of earthquake hypocenters and centroids, using a color-

341    key linearly distributed, piecewise, between 0- and 800-km depths (Table 2).  Subtle differences in color

342    values therefore reflect real depth differences between events.  The depth intervals were chosen consistent

343    with ThERM modes H4, L1, L2, L4, M1, M2, and M4 (Patton and Watkinson, 2010) (Table 3).

344

345    In the upper crust, 0-14.4 km depth, there are about three times as many hypocenters as there are centroids,

346    and the color palette ranges from red to orange (Table 2).  The map distribution of events is similar for both

347    hypocenters and centroids.  Events commonly occur at oceanic ridges, oceanic trenches, and the Alpine-

348    Himalayan belt.  The centroids are distinctly deeper than their corresponding hypocenters.

349

350    In the middle crust, 14.4-25 km depth, the number of hypocenters is only about one-fifth that of centroids

351    and the color palette ranges from orange to yellow.  The map distributions of hypocenters and centroids

352    also show significant differences.  While hypocenters are common at all oceanic trenches and along the

353    Alpine-Himalayan belt, they only rarely appear at oceanic ridges.  Centroids, in addition to occurring at

354    oceanic trenches and the Alpine-Himalayan belt, are common at oceanic ridges and also major transform

355    boundaries, like the San Andreas-Queen Charlotte-Fairweather fault system.

356

357    In the lower crust, 25-37 km depth, hypocenters outnumber centroids by about 1.5 times, and the color

358    palette ranges from yellow to green.  Oceanic trenches are well-populated by both hypocenters and 359 centroids, as is the Alpine-Himalayan belt.  Rare intraplate events of both types also appear.  Interestingly,

360 centroids also sparsely populate oceanic ridges, but primarily at the shallower (yellow) end of the range.

361

362 In the lithosphere, 37-99.5 km depth, the number of hypocenters and centroids is comparable, and the color

363 palette ranges from green to blue.  Hypocenters and centroids are absent at all oceanic ridges and active

364 transform boundaries, but commonly populate all oceanic trenches as well as the Alpine-Himalayan belt.

365 Distinct 'knots' of events appear beneath the Carpathians and the Hindu Kush, with rare events in East

366 Africa-Madagascar.  The map distributions of hypocenters and centroids are similar in most respects.

367

368 The number of hypocenters and centroids in subducting slabs transiting the upper tectosphere, 99.5-172 km

369 depth, is comparable, and the color palette ranges from blue to purple.  As in the preceding depth interval,

370 the map distributions of events are very similar.  Both hypocenters and centroids are well-represented at all

371 oceanic trenches, but also sparsely populate the Alpine-Himalayan belt.  The Carpathian and Hindu Kush

372 'knots' persist to these depths as well.

373

374 The number of hypocenters and centroids in subducting slabs transiting the lower tectosphere, 172-255 km

375 depth, is comparable, and the color palette ranges from purple to pink.  The maps again are very similar,

376 with hypocenters and centroids common at all oceanic trenches.  The Hindu Hush 'knot' persists to these

377 depths as well, but the Carpathian one is absent.  Some events also appear beneath the Aegean and southern

378 Italy.

379

380 The number of hypocenters and centroids in subducting slabs transiting the asthenosphere, 255-690 km

381 depth, is comparable, and the color palette ranges from pink to white.  Again their map distributions are

382 very similar.  Hypocenters and centroids are associated almost exclusively with active subduction margins,

383 e.g. Andes, Tonga-Lau, Malaysia, Indonesia, Japan, Aleutians, South Georgia and Sandwich Islands,

384 except for a few events in the western Alpine Belt, particularly beneath southern Spain and Italy.  The

385 Hindu Kush 'knot' appears pink (shallow) in this depth range, consistent with the base of the tectosphere.

386

387     There are only a handful of events in the mesosphere, where the color palette is uniformly black. These

388     events lie only a few kilometers below 690-km depth. Most appear at the Tonga trench, but a few centroids

389     also lie between Japan and Kamchatka.

390

391     In summary, the depth distribution of earthquakes at various plate margins appears to be more consistent

392     for centroids than it does for hypocenters. Thus, while the latter distribution could be interpreted as

393     indicating significant differences in thickness and mechanical properties between oceanic and continental

394     lithosphere, the former distribution suggests the opposite. In both cases, however, the distribution of

395     earthquakes at convergent margins is distinct from that at divergent and transform margins. It is likely that

396     some of the apparent difference between continents and ocean basins is due to the fact that travel-time

397     models are optimized for continents (Kennett and Engdahl, 1991), where the majority of seismic receivers

398     are located. Nevertheless, there are good rock-mechanical reasons to believe that significant differences in

399     the mechanical properties of continents and the ocean basins exist. Therefore, a detailed study of

400     earthquake depth-energy release from tectonically regionalized data might help quantify the nature of these

401     differences.

402

403     2.3 Regional variations

404     Given differences in the depth distribution of earthquakes, noted above, and their apparent correlation with

405     tectonics, it is natural to consider regional subsets of the CMT catalog. These subsets (Figure 2) over

406     sample the CMT catalog by about 0.8% (Table 4). This is due to the expedient method used to select event

407     subsets, particularly at convergent margins where data density is high. It is unlikely, however, that this

408     small discrepancy has any real impact on the conclusions of this report. A more in-depth analysis of these

409     data, including detailed consideration of CMT dislocation characteristics, is in progress.

410

411     Earthquake depth-energy release curves $\Sigma M_w(z;t)$ present the sum of moment magnitudes $M_w$ for

412     earthquakes, filtered for depth $z$ using a boxcar of selected thickness $t$, at every kilometer from the surface

413     to about 700-km depth. They are an elaboration upon similar curves presented by Frohlich (1989) in his

414     review of deep-focus earthquakes. In early work with the CMT catalog, filter thicknesses of 1, 3, 5, 7, 9,

415 and 11 km, were used, but did not significantly change the depth patterns shown here. The primary effect

416 of filter thickness, apart from curve smoothing, is to change the magnitude of these sums. Consequently,

417 most curves presented here are computed using a 3-km filter, which matches the mean standard depth error

418 of CMTs. This seems to provide adequate smoothing without discarding potential signal. Because moment

419 magnitude is related to scalar seismic moment $M_0$ [dyne-cm] by the formula $M_w = (2/3)log_{10}(M_0)-10.7$ (Aki

420 and Richards, 2002), these curves serve as simple dimensionless proxies for seismic energy release with

421 depth in the planet. How this prevalent, stochastic, and highly localized mode of energy dissipation might

422 be related to energy dissipation in the larger geotectonic system is of primary interest in studying these

423 plots.

424

425 As discussed earlier, well-observed earthquakes with $M_w > 5.0$ are relocated routinely as part of the CMT

426 inversion procedure, using the associated hypocenters as initial estimates. Observed differences between

427 hypocenter and centroid depth-energy release curves can, in part, be understood in this manner. However,

428 the depth-energy release pattern for hypocenters (Figure 3a) is the same for all six tectonic subsets of the

429 earthquakes studied here, as well as for the entire dataset. This is somewhat artificial, and probably reflects

430 optimizations in the quick location algorithms used for event hazard monitoring. On the other hand, the

431 depth-energy release curves for centroids have a more natural appearance (Figure 3b), and reveal

432 interpretable depth structure which apparently depends on the nature of tectonic margins.

433

434 The amplitude of these depth-energy release curves is proportional to the number of events in the given

435 data subset. Consequently, a normalization scheme is needed to enhance possible variations in the depth

436 signal they contain. Given the strong correlation of summed moment magnitude with the number of

437 earthquakes in each subset (Table 4), an expedient normalization scheme is to simply divide $\Sigma M_w(z;t)$ by

438 the number of events $N$ in the subset. Event-normalized earthquake depth-energy release curves

439 $\Sigma M_w(z;t)/N$ reveal significant differences between divergent and transform margins on the one hand and

440 convergent margins on the other. In both panels of Figure 4, the depth-release curve for the entire set of

441 relocated CMTs (black dashed, Table 4) serves as a baseline for these regional comparisons.

442

443   *2.3.1 Divergent and transform margins*

444   The three colored curves shown in Figure 4a correspond to event populations occurring at divergent and

445   transform margins, as plotted in Figure 2a, c, e.  There is relatively little energy release associated with the

446   brittle upper crust.  In the middle crust, from 14-25 km depths, the most seismic energy is dissipated at

447   continental transforms and oceanic ridges, while the least is dissipated at continental rifts.  At all three

448   boundary types, the energy dissipation per event is greater than the global CMT baseline.  It is possible that

449   the greater dissipation of energy at continental transforms, compared to the oceanic crust, is due to the

450   lesser Coulomb strength of typical felsic lithologies found there, when compared to mafic ones.

451

452   In the lower crust, from 25-37 km depths, the most seismic energy is dissipated at continental rifts, at a per-

453   event rate higher than the global baseline.  In contrast, seismic energy dissipation at oceanic ridges and

454   continental transforms are both significantly below the global baseline, with the least dissipation occurring

455   at continental transforms.  This pattern is consistent with the observation that continental crust generally is

456   thicker than ocean crust, and suggests aseismic creep in the lower crust of continental transforms.  The

457   similar amounts of energy dissipation in oceanic and continental crust at extensional margins, combined

458   with their differences in thickness, again suggest that oceanic crustal rocks are stronger than continental

459   rocks.

460

461   Below 37-km depth, all three boundaries dissipate less energy than the global per-event baseline, with the

462   most dissipation occurring in continental rifts, and the least in oceanic ridges.  Below about 50-km depth,

463   the depth-energy release curves disappear altogether (Table 4).  Coincidentally, these depths correspond to

464   a range of pressures thought to be important for the extraction of mid-ocean ridge basalts (MORB)

465   (Presnall and Gudfinnsson, 2008).  Perhaps the fact that basalt volcanism is common at oceanic ridges and

466   continental rifts (e.g. East Africa), and not at continental transforms, reflects decompression melting in the

467   mantle at divergent margins.

468

469   *2.3.2 Convergent margins*

470    The three colored curves shown in Figure 4b correspond to event populations occurring at convergent

471    margins, as shown in Figure 2b, d, f.  Again, there is relatively little energy release associated with the

472    brittle upper crust.  In the middle crust, from 14-25 km depths, the most seismic energy is dissipated at

473    Himalayan-type margins, and at a per-event rate higher than the global CMT baseline.  In contrast, the

474    dissipation occurring at island arc-type and Andean-type margins is below the per-event baseline, with the

475    least dissipation occurring at Andean-type margins.  This might reflect the juxtaposition of generally

476    thicker and weaker continental crust on both sides of Himalayan-type margins, when compared to the other

477    two margins types.

478

479    In the lower crust, from 25-37 km depths, the most seismic energy is dissipated at island arc-type margins,

480    but at rates only slightly greater than the global per-event baseline.  Seismic energy dissipation at Andean-

481    type and Himalayan-type margins is below the global baseline, and while the Andean-type release curve

482    closely parallels the baseline, the Himalayan-type curve deviates substantially.

483

484    Below 37-km depth, seismic energy dissipation in Himalayan-type margins is noticeably less than the

485    global baseline to about 75-km depth.  Earthquake depth-energy release curves for convergent margins

486    persist to depths of 492 km at Himalayan-type margins, 656 km at Andean-type margins, and 699 km at

487    island arc-type margins (Table 4).  The apparent contrast between convergent and divergent/transform

488    margins, earlier discerned from the seismicity maps, is clearly reflected in these depth variations.

489

490    *2.3.3 Variations throughout the asthenosphere*

491    Given the great depth to which seismic activity occurs at convergent margins, it makes sense to examine

492    deviations from the global baseline throughout the asthenosphere.  Again, the depth-release curve for the

493    entire set of relocated CMTs (black dashed, Table 4) serves as a baseline for these regional comparisons.

494    Figure 5 shows event-normalized release curves for the three convergent margin types, filtered using a 10-

495    km boxcar.  Relatively large deviations from the global baseline are apparent, particularly at depths less

496    than about 260 km.

497

498    In the lithosphere, from 37-100 km depths, seismic energy dissipation at Himalayan-type margins is less

499    than the baseline to about 75-km depth, as noted earlier, while that at Andean-type and island arc-type

500    margins closely parallel the baseline.  This might reflect the fact that subduction of oceanic lithosphere is

501    ongoing at the two latter margins, but has ceased at the former.

502

503    In the upper tectosphere, from 100-175 km depths, the greatest energy dissipation occurs at Andean-type

504    margins, and at rates higher than the global baseline.  Through this range, energy dissipation at Himalayan-

505    type and island arc-type margins closely tracks the global baseline, except at about 160-180 km depth,

506    where dissipation in the Himalayan-type margins falls off.  Coincidentally, these depths correspond to the

507    range of pressures for which peridotite xenoliths in kimberlites show distinct planar tectonite fabrics (Boyd,

508    1973; James et al., 2004).  Perhaps this reflects localized weakening of the mantle at these depths,

509    consistent with isobaric shearing at mode M1 of ThERM (Patton and Watkinson, 2010) (Table 3).

510

511    In the lower tectosphere, from 175-260 km depths, seismic energy dissipation at Himalayan-type and

512    Andean-type margins noticeably differ from the global baseline, while that at island arc-type margins

513    closely tracks the baseline.  This is consistent with the presence of continental tectosphere (Jordan, 1975) at

514    the former two margin types, and its absence at island arc margins.

515

516    In the asthenosphere, from about 500-650 km depths, the rate of seismic energy dissipation increases.  This

517    is a depth range where mineral phase transformations are thought to be likely (Birch, 1952; Ringwood,

518    1991), and it is possible that these earthquakes represent a phase-transformation 'anti-crack' population

519    (Green, 2005), although there are other hypotheses (Frohlich, 1989).  Seismic energy dissipation tails off

520    substantially at about 660-km depth, before disappearing altogether at about 700-km depth.  This cut-off of

521    seismicity roughly corresponds to modes M3 and M4 of ThERM (Patton and Watkinson, 2010).

522

523    In summary, many, but certainly not all, of the variations of earthquake depth-energy release from the CMT

524    catalog correspond with the boundary layer structure of ThERM (Patton and Watkinson, 2009, 2010)

525    (Table 3).  The coincidence of inflections, triplications, and minima in these release curves with the

526    predicted depths of isobaric shearing modes suggests a coherent layering of thermomechanical competence

527    with depth in the planet.  Furthermore, the marked differences in event-normalized energy release between

528    tectonic regions suggest significant lateral variations.  Consequently, it makes sense to explore further the

529    implications of this model for dynamic plate theory.

530

531    **3.0 Thermodynamics of solid self-gravitating matter configurations**

532    <u>3.1 Preliminaries</u>

533    The macroscopic notion of heat is defined as the difference between the internal energy and work

534    performed on a system, consistent with the Joule heating experiments (Chandrasekhar, 1967).  The First

535    Law of thermodynamics is therefore

$$dQ = dU - dW \qquad (2)$$

537    where $dQ$, $dU$ and $dW$ are increments of heat, internal energy and work, respectively.  Note that heat is a

538    derived quantity, having no meaning independent of the First Law.

539

540    Guided by triaxial rock mechanics experiments, I shall account for the entropy density of strained solid

541    materials using a simple one-dimensional elastic model.  This model exhibits an unorthodox behavior

542    consistent with Lavenda's (1995) notion of thermodynamic symmetry breaking.  I deduce the expected

543    slope, shape, and temperature dependence of the energy density for this model, which then serves as foil for

544    the thermomechanics of shear localization exhibited by non-linear elastic DG-2 materials.  Note that while

545    entropy appears in all statements of the Second Law, it was first formulated axiomatically by Carathéodory

546    (ca. 1909), based on an analysis of Pfaffian differential equations, to read

$$dQ = TdS \qquad (3)$$

548    where $dS$ is an increment in the entropy density, and $T$ is absolute temperature.

549

550    Consider a cylindrical test specimen of rock, with length $l$ and diameter $d$, placed in a loading frame for the

551    purpose of strength characterization.  Upon applying a force $\varphi$ directed along a line parallel to the

552    specimen's length, it is observed that the specimen shortens by a length increment $dl$.  Consequently, the

553    increment of work needed to shorten the cylinder from $l+dl$ to $l$ is given by

554
$$dW = \varphi dl \,. \tag{4}$$

555 Because the specimen can be held under relatively small loads for long periods of time, it is reasonable to

556 assume that it manifests a force equal and opposite to the applied load.  Presumably, this reaction force

557 arises from electromagnetic interactions in the sample's microstructure.  Furthermore, experience shows

558 that if the cylinder were unloaded, it would likely return to its original length.  This is Hooke's law (*Ut*

559 *tensio sic vis*, ca. 1642).

560

561 However, it is equally valid to consider this problem from a material point of view.  Responding to a

562 directed environmental load of magnitude $\varphi$, the cylinder strains by an increment $dl$ of its overall length $l$,

563 and as a result distributes an increment of energy $dU$ throughout its microstructure and mineral fabric.

564 Therefore, I am also free to assume a macroscopic relation of the form

565
$$dU = \varphi dl \,. \tag{5}$$

566

567 From a thermodynamic point of view, the unloaded state to which the cylinder tends to return upon

568 unloading is somehow more likely than the loaded one, and should therefore coincide with a maximum in

569 entropy.  Consequently, any deformation of the cylinder from this ideal state must necessarily decrease the

570 entropy of the cylinder itself, as a function of length.  Consequently, I seek a relation of the form

571
$$dS = -f(l)dl \,. \tag{6}$$

572 Combining the First and Second Laws, equations (2) and (3), I obtain

573
$$TdS = dU - dW \,. \tag{7}$$

574 Upon substituting for the work and internal energy increments using (4) and (5) respectively, I find

575
$$TdS = \varphi dl - \varphi dl \Rightarrow dS = 0 \tag{8}$$

576 Hence, for a positive absolute temperature thermodynamics predicts no increase in entropy, consistent with

577 the apparent lack of energy dissipation, a state of mechanical equilibrium for cylinders under small loads,

578 and everyday experience.

579

580

581    3.2 Strained inhomogeneous elastic solids

582    Curiously, there is no need to account for heat in these experiments, despite its fundamental importance in

583    thermodynamics. It would appear that there is no meaningful distinction to be made between heat and

584    work for this model (Lavenda, 1995). Instead the thermodynamic potentials for work, internal energy, and

585    entropy are all functions of a single variable. This has immediate consequences for the usual combination

586    of the First and Second Laws, equation (7). Because heat and work are indistinguishable, and heat is

587    already accounted for in the product of the temperature and the entropy increment through the Second Law,

588    equation (7) can be rewritten as

589
$$\frac{f(l)}{\varphi} = \frac{1}{T} = -\frac{d\Delta S}{d\Delta U} .$$
(9)

590    Here the entropy and internal energy are prefixed with deltas to distinguish these *primitive functions* from

591    the classical thermodynamic potentials assumed above, which are first-order homogeneous functions.

592    These primitive functions are *inhomogeneous*, which is to say that they can manifest scale-dependence,

593    contrary to the scalability expected of classical homogeneous potentials. In the following section I explore

594    the scale-dependence of this system, by employing power laws for initial statistical distributions in length.

595

596    The importance of statistical variability in the behavior of elastic materials, and rocks in particular, is

597    demonstrated by modeling the entropy and energy potentials for this system as power laws in length $l$.

598    Following Lavenda (1995), I define the *internal energy increase* as

599
$$\Delta U(l) = \frac{\eta}{m} l^m$$
(10)

600    and the *entropy reduction* as

601
$$\Delta S(l) = -\frac{k\sigma}{n} l^n$$
(11)

602    where $k$ is Boltzmann's constant, $\sigma$ and $\eta$ are positive constants independent of temperature, and $n$ and $m$

603    are positive numbers.

604

605    Observe that equation (9) can be rearranged to represent the temperature as

606
$$T = -\frac{d\Delta U}{d\Delta S} . \tag{12}$$

607 Upon differentiating the primitive functions (10) and (11) with respect to length, and substituting into

608 equation (12) I obtain

609
$$T = \frac{\eta}{k\sigma} l^{m-n} . \tag{13}$$

610 Consequently, the temperature of this model system can either increase or decrease with length, depending

611 on whether the exponent $m$ - $n$ is positive or negative. The temperature is independent of length for $m = n$.

612

613 The modulus of elasticity $E$ for the model is defined by the derivative of the force $\varphi$, which in turn is the

614 derivative of the internal energy, equation (5). Upon eliminating length in this expression via the

615 temperature relation, equation (13), I obtain

616
$$E = (m-1)\eta \left( \frac{k\sigma T}{\eta} \right)^{\frac{m-2}{m-n}} . \tag{14}$$

617 For $n = 2$, the modulus of elasticity is $E=(m-1)k\sigma T$, and the force reduces to a generalized Hooke's law

618 $\varphi=E(T)l$. This relation further reduces to a linear force-displacement law, but only when the internal

619 energy too is quadratic in length, $m = 2$.

620

621 Upon inverting equation (13) to express length as a function of temperature, differentiating the result with

622 respect to temperature, and eliminating the constants via (13) I obtain

623
$$\frac{1}{l}\frac{dl}{dT} = \frac{1}{(m-n)T} . \tag{15}$$

624 This expression characterizes the thermal elongation of the model, analogous to the thermal expansivity in

625 three-dimensions. The latter material property dictates the scale over which body forces can act in the

626 model. Consequently the model elongates upon heating for $m > n$, shortens upon heating for $m < n$, and is

627 undefined for $m = n$.

628

629     By separating the entropy and energy increments in equation (12), dividing through by a temperature

630     increment $dT$, and employing the chain rule to express the common length dependencies for entropy,

631     energy, and temperature, I obtain

632
$$\frac{d\Delta U}{dl}\frac{dl}{dT} = -T\frac{d\Delta S}{dl}\frac{dl}{dT} \ . \tag{16}$$

633     Upon substituting the derivatives of equations (10), (11), and (13) with respect to length into (16) I find

634
$$\frac{d\Delta U}{dT} = -T\frac{d\Delta S}{dT} = \left(\frac{\eta}{m-n}\right)\frac{l^m}{T} \tag{17}$$

635     For positive absolute temperature, this shows that the heat capacity of this inhomogeneous elastic system

636     cannot be defined simultaneously as

637
$$C \equiv \frac{dQ}{dT} = T\frac{d\Delta S}{dT} \tag{18}$$

638     and

639
$$C \equiv \frac{dQ}{dT} = \frac{d\Delta U}{dT} \tag{19}$$

640     because one of these definitions always will be negative when the other is positive, and *vice versa*. Apart

641     from the pathological case for $m = n$, there are two other distinct types of inhomogeneous elastic systems

642     depending on whether $m < n$ or $m > n$.

643

644     Variability as a function of length is inversely proportional to the exponent appearing in the primitive

645     power laws, above. Hence, a smaller exponent means greater variability. Equation (10), defining the

646     internal energy increase, is associated with *mechanical* variability. However, equation (11), defining the

647     entropy reduction, is associated with *statistical* variability rather than thermal variability, because heat is

648     not evident in this problem. Consequently, systems dominated by either mechanical variability ($m < n$) or

649     statistical variability ($m > n$) can be identified on the basis of their heat capacity $C$, as follows:

650
$$C \equiv \frac{dQ}{dT} = \begin{Bmatrix} T\dfrac{dS}{dT}, m < n \\[2mm] \dfrac{dU}{dT}, m > n \end{Bmatrix} \tag{20}$$

651   These conclusions are easily substantiated by returning to thermodynamic fundamentals. Eliminating

652   length between the primitive equations (10) and (11), I find for the case of mechanical variability that

$$\Delta S \sim -\left(\Delta U\right)^{\frac{n}{m}}. \tag{21}$$

654   In words, the *entropy* is a *concave function* of the internal energy. Also, for the case of statistical

655   variability I find that

$$\Delta U \sim \left(|\Delta S|\right)^{\frac{m}{n}}. \tag{22}$$

657   In words, the *internal energy* is a *convex function* of the entropy. Furthermore, because *dS/dU < 0*, these

658   representations are mutually exclusive. The usual symmetry of the entropy and energy representations,

659   expected from equilibrium thermodynamics and arising from first-order homogeneity of thermodynamic

660   potentials, is broken.

661

662   For an inhomogeneous elastic system dominated by *mechanical variability* (*m < n*), *dQ* is the amount of

663   heat *evolved* by the system, which leads to a decrease in entropy by an amount *dS=dQ/T* (Figure 6a).

664   Therefore fewer microscopic states are available at lower temperatures. The slope of the concave entropy

665   density function is -1/T; higher temperatures are associated with flatter slopes, and lower temperatures with

666   steeper slopes. This model elongates upon cooling. Because temperature and heat are both decreasing

667   functions of length, the entropy density is proportional to length. Mechanical variability offers no insight

668   for the thermodynamics of DG-2 materials.

669

670   On the other hand, in an inhomogeneous elastic system dominated by *statistical variability* (*m > n*), *dQ* is

671   the amount of heat *absorbed* by the system, which leads to an increase in internal energy by an amount *dU*

672   *= dQ > 0* (Fig. 6b). This corresponds to a decrease in the entropy by an amount *dS = -dQ/T*. Consequently

673   there are fewer microscopic states available at higher temperatures. The slope of the convex energy density

674   function is –*T*; higher temperatures are associated with steeper slopes, and lower temperatures with flatter

675   ones. This model elongates upon heating. Furthermore, because temperature and heat are both increasing

676   functions of length, energy density is inversely proportional to length. Statistical variability offers crucial

677   insights for the thermodynamics of DG-2 materials.

678

679 Nowhere in this simple one-dimensional model has the phenomenon of shear failure been addressed.  For

680 example, if we repeatedly load our test specimen, or apply progressively higher loads, experience tells us

681 that the specimen will, at some point, spontaneously fail, sometimes after suffering significant

682 microphysical damage (Katz and Reches, 2004).  In other words, the simple act of loading the test cylinder

683 changes its prior microstructure and mineral fabric.  Although the detailed distribution of these changes

684 cannot be known to an outside observer, they can be treated statistically, as was appreciated by Weibull (ca.

685 1939).  These issues are addressed in Section 4.0.

686

687 3.3 Self-gravitating matter configurations

688 Lavenda (1995) shows that the thermodynamics of a self-gravitating body, like a planet or star, is also

689 subject to symmetry breaking of the type outlined above.  In this case, like that of inhomogeneous elastic

690 systems dominated by statistical variability, the energy density is given by a monotonically decreasing

691 function of the entropy density (Figure 6b).  Significantly, the energy density is inversely proportional to

692 the body's radius.  In other words, both pressure and temperature increase with depth.

693

694 The heat capacity of a self-gravitating body is defined by

695
$$C \equiv \frac{dQ}{dT} = \frac{dU}{dT} = \frac{mc^2}{T_0} \qquad (29)$$

696 where $m$ is the mass of an elementary particle, $c$ is the speed of light, and $T_0$ is a reference temperature.

697 The energy density, and by association the stress, of a self-gravitating body plays a fundamental role in

698 shaping the body, and in controlling its spacetime evolution, in accordance with the far reaching

699 implications of Einstein's (1916) theory of gravitation.

700

701 **4.0 Thermomechanics of DG-2 materials**

702 4.1 Diharmonic equation

703 The pure-shearing plane-strain deformation of non-linear elastic DG-2 materials is governed by the

704 diharmonic equation (Patton, 1997)

705
$$0 = \alpha^2 \frac{\partial^4 \psi}{\partial x^4} + \left(1 + \alpha^2\right)\frac{\partial^4 \psi}{\partial x^2 \partial z^2} + \frac{\partial^4 \psi}{\partial z^4}$$
(24a)

706
$$\alpha^2 = \left(1 - 2(\kappa/\chi)\right)/\left(1 + 2(\kappa/\chi)\right)$$
(24b)

707 where the ratio of thermal $\kappa$ to mechanical $\chi$ diffusivities is called thermomechanical competence. Note

708 that as $\kappa/\chi \rightarrow 0$, equation (24) reduces to the biharmonic equation, which appears in theories of linear

709 elastic and linear viscous materials.

710

711 4.2 Incipient modes analysis

712 Here incipient modes analysis (Patton and Watkinson, 2010) is used to document the deformation modes of

713 DG-2 materials and thereby facilitate comparison with the thermodynamic properties of simple

714 inhomogeneous elastic materials, documented in section 3. I substitute wave-like harmonic and

715 dislocation-like shear band solutions into the differential equation (24) to identify domains were

716 deformation modes of these types are possible. In all cases, domains of thermomechanical competence that

717 allow real roots, or at least roots with real parts, will admit solutions of the assumed type. In both trial

718 solutions $\psi$ is the stress-energy function, identically satisfying incompressibility. This linear analysis in no

719 way constrains the finite growth of the resulting structures.

720

721 For these analyses, consider a two-dimensional spatial domain in which a specific but arbitrary set of

722 Cartesian axes are drawn through an arbitrarily chosen point. Consider also a plane strain deformation field

723 where the velocity components *(u, w)* expressed in this coordinate system are assumed proportional to the

724 distance from the origin in the following way, *(u, w) ∝ (-x, z)*.

725

726 The usual Cartesian harmonic normal modes

727
$$\psi = \exp^{i\omega x + rz}$$
(25)

728 suffice for the wave-like case. Here a relatively competent layer of thickness $H^*$ has its mean position

729 parallel to the shortening direction, and it is initially planar. In other words, the boundaries between the

730 layer and the weaker matrix initially lie at $z = \pm H^*/2$. As the layer shortens in the base field, it tends to

731 thicken, and harmonic perturbations with amplitude $\delta^*$ might begin to develop. The growth of such

732 undulations into observable folds or waves depends on the relative rates of their amplification versus

733 uniform layer thickening. The wavenumber $\omega$ predicts the normalized wavelength of the incipient

734 undulation through the relation $L^*/H^* = 2\pi/\omega$. The small scalar parameter in this case is $\varepsilon = \delta^*/H^*$, so that

735 these deductions are rigorous only for the case of infinitesimal fold amplitude.

736

737 Substituting the harmonic trial solution (25) into (24a) I obtain the four distinct roots

738
$$r = \pm\omega \quad \text{or} \quad r = \pm\alpha\omega \; . \qquad (26)$$

739 These are all real for $0 < \kappa/\chi < \frac{1}{2}$, thus wave-like material deformations are expected to form on this

740 domain of relatively low thermomechanical competence, in response to far-field forcing (Figure 7a, solid

741 blue curve). Note also that for $\frac{1}{2} < \kappa/\chi < \infty$ these roots are mixed, with two real and two pure imaginary, so

742 that harmonic disturbances are also possible on this domain of higher thermomechanical competence

743 (Figure 7a, dotted blue curve). Together, these harmonic perturbations represent shear waves (reversible

744 rotational distortions) propagating throughout the material (Truesdell, 1964). Furthermore, given the

745 planetary scale implications of DG-2 thermodynamics, material coupling on the domain of lower

746 competence could explain some aspects of long-period Love wave attenuation in the upper mantle (Table

747 5). The normalized wavelength of these disturbances scales as $L^*/H^* = (\kappa/\chi)^{-1}$ (Patton and Watkinson,

748 2005).

749

750 In the dislocation-like case, I use shear band solutions

751
$$\psi = \exp^{g_z x - g_x z} \qquad (27)$$

752 following the work of Hill & Hutchinson (1975) and Needleman (1979). The vector $g$, with components $g_x$

753 and $g_z$ in the chosen coordinate system, is normal to the incipient shear band. Consequently, the arctangent

754 of the (real) ratio $g_x/g_z$ determines the angle between the band itself and the shortening direction. The ratio

755 of the thickness of the incipient band $\delta^*$, to its length $L^*$, provides a suitable small scalar parameter $\varepsilon$, so

756 that these deductions are rigorous only for the case of a vanishingly thin band.

757

758 Substituting the shear band trial solution (27) into (24a), I obtain the four distinct roots

759
$$\frac{g_x}{g_z} = \pm i \quad \text{or} \quad \frac{g_x}{g_z} = \pm i\alpha \; . \qquad (28)$$

760 While the first pair of roots is purely imaginary, the second pair is real for $\frac{1}{2} < \kappa/\chi < \infty$, where the rescaling

761 modulus (24b) itself is imaginary. Consequently, dislocation-like disturbances are expected to form at

762 angles, symmetric about the loading axis, ranging from 0 to 45 degrees for these relatively high values of

763 thermomechanical competence (Fig. 7a, red curves). Significantly, an angle of 30 degrees corresponds

764 with the value $\kappa/\chi = 1$, which defines the lower limit of thermomechanical rigidity. This suggests that

765 dislocation-like disturbances should form, preferentially, at about 30 degrees with respect to the shortening

766 direction in thermomechanically competent materials (Patton and Watkinson, 2011; Patton and Watkinson,

767 in review). This prediction is consistent with Anderson's 1905 theory of crustal faulting (Jordan et al.,

768 2003).

769

770 4.3 Stress-energy density map

771 The thermomechanics of DG-2 materials can be represented graphically (Patton, 2005; Patton and

772 Watkinson, 2005) (Figure 7b) using three energy thresholds, one for distributed harmonic deformations

773 (green curve, $\psi^D = (\kappa/\chi)^{-1}$), one for intrinsic strain-energy storage (gray curve, $\psi^I = (|\alpha|)^{-1}$), and another for

774 localized shearing deformations (orange curve, $\psi^L = (|\alpha|\kappa/\chi)^{-1}$). Taken together, these curves define a

775 statistically stable stress-energy density map for these materials. Observe that the three threshold curves

776 are monotonically decreasing on certain domains of thermomechanical competence, and that the stress-

777 energy density approaches infinity ("blows up") at the lower end of these respective domains. The

778 distributed threshold curve has a vertical asymptote at $\kappa/\chi = 0$, while the intrinsic curve has one at $\kappa/\chi = \frac{1}{2}$.

779 The localization threshold curve, defined as the product of the other two, consequently exhibits two

780 asymptotes, with a distinct non-zero energy minimum between them. The presence of two distinct energy

781 spikes in this diagram, and their diffusive connection via the dynamic rescaling theorem (Patton and

782 Watkinson, 2010), give rise to all of the geologically interesting behavior of the DG-2 material.

783

784 **5.0 Application to rock mechanics**

785    The foregoing analysis can be applied to the interpretation of data from rock mechanics experiments on

786    cylindrical specimens.  Three facts make this possible.  First, the earlier analysis of statistical variability

787    (Figure 6b) shows that the energy density for strained elastic solids must be a monotonically decreasing

788    function of the entropy, which in turn must be an inverse function of length.  Thus, the diameter/length

789    ratio ($d/l$) of a cylindrical specimen serves as the abscissa for plotting these data (Table 6).  Second,

790    because the intrinsic stress-energy density threshold ($\psi^I$, Figure 7b) arises from the differential normal-

791    stress term of the DG-2 constitutive equation (Patton, 1997; Patton and Watkinson, 2005; Patton and

792    Watkinson, 2010; Patton and Watkinson, 2011) it makes sense to plot normalized differential stress (NDS)

793    from the experiments as the ordinate.  Using Katz & Reches' (2002, 2004) definition of *NDS = ($\sigma_1$- $\sigma_3$)/586*

794    and *their* identification of *NDS ~ 0.96* as a critical threshold for spontaneous sample failure, it is a simple

795    matter to see that the number *$\zeta = 4\sqrt{3}$*, common in the theory of DG-2 materials (Patton and Watkinson,

796    2010), reconciles their conclusion with this thermomechanical theory.  The final and seemingly obvious

797    fact is that the deformation of any rock sample in any load frame is possible only because of the

798    *thermomechanical rigidity* of the loading frame itself.  Thus, the deformed samples and the loading frame

799    comprise complementary parts of a larger, and more interesting, thermomechanical system.  The samples

800    plot at *($\kappa/\chi$ ,$\psi$) = (d/l, $\zeta$ ($\sigma_1$- $\sigma_3$)/586)*, while the loading frame plots at *($\kappa/\chi$ ,$\psi$) = (1, 1)*.

801

802    Fourteen samples of the medium-grained Mt. Scott granite were loaded to predetermined values of axial

803    stress $\sigma_1$, at confining pressure $\sigma_3$ = 41 MPa, and held for predetermined periods, before the loads were

804    slowly released.  Twelve samples that did not suffer macroscopic failure are plotted as green circles (Figure

805    7b).  Three of these samples (105, 124, & 125, Figure 7b, inset) were reloaded to failure at higher loads, as

806    indicated by gold diamonds and thin dashed lines.  Three other samples loaded to failure also are plotted as

807    gold diamonds.  Three samples (#104, # 106 and #110), plotted as red diamonds, spontaneously failed

808    during their hold periods.  Given the standard error in measured Coulomb strength for the Mt. Scott granite

809    (*586 ± 16 MPa*), indicated by the thin dashed lines plotted above and below the distributed and localized

810    threshold curves (inset), all the failed samples plot above the localization threshold.

811

812    Katz & Reches (2002, 2004) observed the microscopic and macroscopic damage to several samples after

813    loading.  They report two populations of microscopic cracks.  Population A are intragranular tension cracks

814    (Figure 7a, red dashed curve) with angles, in relation to the loading direction, in the range 0º – 10º, while

815    Population B are intergranular shear fractures (Figure 7a, red solid curve) with angles in the range 11º –

816    40º.  They also report angles of macroscopic shear failures, which tend to increase with confining pressure.

817    The macroscopic shear angles correspond to thermomechanical competences in the range $0.6 < \kappa/\chi < 0.9$.

818    These values are consistent with 'internal friction' values for Anderson-type brittle faulting.  Additionally,

819    the angle of the macroscopic shear developed in specimen #110, which suffered spontaneous failure during

820    its hold period, can be connected to its pre-failure loading state by a tie line LB-LY', tangent to the

821    localization curve.  Thus, the macroscopic failures observed in these triaxial loading experiments appear to

822    be consistent with the predictions of dynamic rescaling (Patton and Watkinson, 2010).  In contrast, the

823    microscopic Population B shear fractures correspond to the range $0.54 < \kappa/\chi < 2.9$.  This spread might

824    reflect the inhomogeneous distribution of lengths in the samples prior to loading, the effect of confining

825    pressure, or both.

826

827    Data from the three spontaneously failed samples can be used to estimate the mechanical diffusivity as $\chi =$

828    $l^2/\tau$.  The resulting values fall in the range $2.3 \leq$ -$log\ \chi \leq 5.6$.  These estimates are in or below the range $–log$

829    $\chi \leq 5.3$-$5.7$, predicted by theory (Patton and Watkinson, 2010), and overlap with ranges estimated from

830    GPS strain rates, $4.3 \leq –log\ \chi \leq 7$, GPS differential velocities, $4.4 \leq –log\ \chi \leq 5.2$, and structural data, $4.2 \leq –$

831    $log\ \chi \leq 4.7$.  This correlation suggests that combined analysis of data from experimental rock mechanics

832    and GPS surveys will factor highly in further development of dynamic plate theory.

833

834    **6.0 Isobaric shearing hypothesis**

835    In a self-gravitating fluid body, pressure $p$ increases with depth according to the hydrostatic relation $p =$

836    $\rho gz$, where $\rho$ is mass density, $g$ is gravitational acceleration, and $z$ is depth.  Similarly, pressure increases

837    with depth in a solid body according to a lithostatic relation, given by the tensor trace of its stress-energy

838    density.  The depth to which differential stresses can persist in any planet, then, depends on the definition

839    of its stress-energy density.  However, because even terrestrial planets are spheroidally-shaped, which

840 reflects the combined effects of their self-gravity and rotational momentum, the magnitude of these

841 differential stresses must be small.  Furthermore, given that temperature increases with depth (Figure 6b),

842 and that microphysical mechanisms for solid-state creep are thermally activated, even these small

843 differential stresses must diminish rapidly with depth.  In the static case, they decay entirely, and the

844 pressure becomes for all intents and purposes hydrostatic.  Consequently, the maintenance of differential

845 stresses at depth in a planet requires some dynamic process (McKenzie, 1967).  For more than 40 years,

846 this process has been assumed to conform to the assumptions of the standard Earth model (Bercovici et al.,

847 2000; Bunge et al., 1997; Tackley, 2000).  However, based on the foregoing thermomechanical analysis, it

848 is likely that geodynamics is much more interesting than heretofore recognized.

849

850 With pressure and temperature as boundary conditions on a self-gravitating planet, and confining pressures

851 much larger than potential differential stresses, it is hard to argue that material strength matters, except for

852 the fact that the *dynamic rescaling theorem* (Patton and Watkinson, 2010) focuses deformation on the

853 smallest crystalline structures of a solid system.  This is necessary, so that the global dissipation of energy

854 is minimized.  An immediate consequence of this prediction is that shear waves can be propagated

855 throughout a thermomechanical mantle, whereas in a viscous one, no such propagation is possible.  Absent

856 this theorem, however, one must accept the geodynamicist's approximation, that over long time and length

857 scales the mantle is effectively viscous.  As plausible as this might sound, it is impossible to falsify.

858 Moreover, it is inconsistent with the fact that rocks loaded in the laboratory exist in the solid-state.  Thus, it

859 is clear that the failure of the standard Earth model arises solely from a theoretical deficiency.  On the other

860 hand, with the dynamic rescaling theorem, the only substantive differences between deformation of a rock

861 sample in the laboratory, and tectonic deformation of the Earth, are the relative magnitude of the confining

862 pressure and effect of global conservation laws.  In other words, shear localization at the planetary-scale

863 must account for the incompatibility of rectilinear motions with the spheroidally curved geometry of the

864 planet itself, while in the laboratory this is of no concern.

865

866 The statistically stable thermomechanics of non-linear elastic DG-2 materials are not altered by pressure.

867 Consequently, even under extreme pressures the locus of its three energy thresholds and documented

868    scaling relationships can be expected to hold. Therefore, a thermomechanical Earth model can be formed

869    simply by scaling up to Earth's radial structure and applying pressure and temperature boundary conditions

870    at its surface (Patton and Watkinson, 2009, 2010). The result (Figure 8) immediately predicts a variation of

871    pressure and temperature expected for Earth's mantle. Furthermore, it predicts that the outer colder parts of

872    the mantle should be thermomechanically rigid, thermodynamically isothermal, and subject to brittle shear

873    localization, while the deeper hotter parts should be thermomechanically ductile and thermodynamically

874    adiabatic. Adiabaticity prevails as $\kappa/\chi \rightarrow 0$, consistent with depths in the lower mantle, and coincidentally

875    where Birch (1952) showed it to pertain on the basis of seismic wave speed variations (Equation 1). The

876    asthenosphere of a thermomechanical Earth is not adiabatic, because differential stresses (normal stress

877    differences) there are large enough to cause shear localization. Moreover, the vanishing of seismicity at the

878    asthenosphere-mesosphere boundary, ~700 km deep, reflects this fundamental change in thermodynamic

879    conditions. For comparison, the variation in earthquake depth-energy release (Figure 5) is plotted in the

880    last panel of Figure 8, where the most significant deviations from the per-event global baseline clearly

881    correlate with the low velocity zone, and depths consistent with Earth's lithosphere and tectosphere.

882

883    This model, being self-similar or fractal, can also be used to correlate crustal and upper mantle

884    observations. Figure 9 depicts 'crustal overtones' of ThERM. For comparison, the variation in earthquake

885    depth-energy release for all six margin types considered in this paper (Figure 4) are plotted in the last panel

886    of Figure 9, where the most significant deviations from the global per-event baseline correlate to the

887    seismic lid and crust. What does this remarkable correlation mean for dynamic plate theory?

888

889    For a self-gravitating solid elastic body, like a terrestrial planet, we can anticipate some degree of interplay

890    between the inhomogeneous statistical distribution of length scales in the body, and the distribution of

891    thermal and compositional lengths over which the body force of gravity might act. This interplay is

892    expressed particularly in the structure of the thermomechanical boundary layer that forms adjacent to the

893    cold surface of the planet, but also by the fact that elastic shear waves are propagated throughout the

894    mantle. Thus, in order for any portion of a thermomechanical planet to suffer deformation, there must be

895    measurable contrasts in material competence, as well as concentrated body forces. Furthermore, the 896   mechanical diffusivity $\chi$ must be greater than the thermal diffusivity $\kappa$ in deforming portions of this

897   complex system. Consequently, the thermomechanical boundary layer that forms will always be $\zeta$ times

898   thicker than the purely thermal one. Depth variations in seismic energy release on such a planet are

899   therefore to be expected. Finally, given that pressure and temperature variations are explicitly predicted by

900   theory, it is reasonable to suppose that variations in material competence will exhibit strong dependencies

901   on bulk composition and volatile content.

902

903   In summary, the energy density function for a self-gravitating, inhomogeneous elastic body, dominated by

904   statistical variability (m > n), must be a monotonically decreasing convex function of the radial coordinate

905   and the entropy reduction. Furthermore, the entropy reduction itself is an inverse function of length. Heat

906   absorbed by such a system will tend to increase the internal energy, but correspondingly decrease the

907   entropy. Such a body will not readily evolve heat, except when regional conditions favor a return to more

908   classical thermodynamics. In these subsystems, the energy density function would necessarily exhibit a

909   positive slope. The monotonically increasing branches of the intrinsic and localization thresholds (dashed

910   curves, Figure 7b) exhibit these characteristics, and also correlate with depths in the asthenosphere where

911   magma generation generally is thought to take place (Figures 8 and 9). Magmatic differentiation provides a

912   mechanism for generating density contrasts between continental and oceanic crust, the tectosphere, and

913   average mantle.

914

915   These considerations are quite general and place, once and for all, the stress-energy density thresholds of

916   DG-2 materials (Patton and Watkinson, 2005) in a coherent thermodynamic context. Consequently, the

917   behavior of these ideal materials can be correlated with the pressure, temperature, age, and geometry of

918   geological structures observed in outcrops, orogens, and terrestrial planets. For example, the outer parts of

919   such planets are predicted to be relatively cold, competent, and subject to dynamic shear localization

920   ("brittle"), while the inner parts are predicted to be relatively hot, incompetent ("ductile"), and structurally

921   simple. For Earth, this is reflected in the remarkable correlation of spherically symmetric elastic models,

922   like PREM (Dziewonski and Anderson, 1981), with the predicted depth distribution of isobaric shears in a

923   body with a ~100-km thick lithosphere (Patton, 2001; Patton and Watkinson, 2009, 2010). As 924  demonstrated here, this correlation also holds for variations in earthquake depth-energy release (Figures 3-

925  5).  The following section discusses the implications of these findings for the interpretation of various

926  geophysical data sets, and dynamic plate theory in general.

927

928  **7.0 Discussion**

929  <u>7.1 Laboratory measured versus theoretical viscosity</u>

930  Spontaneous shear dislocations are characteristic modes of deformation for solids under loading, which are

931  related non-linearly to the strength of the loaded material.  In a loading frame in the laboratory, rock

932  strength can be reduced to a steady-state rate of energy dissipation in shear, i.e. 'viscosity'.  This is possible

933  only because the loading frame itself is effectively rigid ('stiff') by comparison.  But workers conducting

934  these experiments, and theoreticians interpreting their results for geodynamic models, surely understand

935  that the materials involved are solids.  So why is the standard Earth model based on viscous fluid

936  dynamics?

937

938  Another important macroscopic property of rocks is density, which is related to chemical composition and

939  mineralogy via specific gravity.  Perhaps, then, a better measure of rock 'strength' is the ratio of a rock's

940  'viscosity' to its density.  This ratio has dimension $L^2 T^1$, which has been called 'diffusivity'.  This latter

941  term is descriptive, in that energy is being dissipated by the imposed deformation, and unburdened by the

942  fluid mechanical connotations of the term 'viscosity'.

943

944  For example, the rate of energy dissipation in the shearing of a rheological fluid depends on the ratio of its

945  molecular viscosity ($\mu$) and its density ($\rho$).  Although we might call this quantity the 'shear stress

946  diffusivity', one usually hears the term *kinematic viscosity* ($v = \mu/\rho$).

947

948  Similarly, the rate of heat dissipation in a substance can be expressed as the ratio of its thermal conductivity

949  ($k$) to the product of its heat capacity ($C_p$) and density ($\rho$), also having dimension $L^2 T^1$.  This quantity is

950  usually called *thermal diffusivity* ($\kappa = k/\rho C_p$), but perhaps some would prefer the term 'thermal viscosity'?

951

952 Finally, the rate of energy dissipation in a deforming rheological solid can be expressed as the ratio of its

953 'viscosity' to its density, which might also be called 'kinematic viscosity'. However, because this brings to

954 mind many ideas that have naught to do with deforming solids, it does little to dispel the lexical confusion I

955 am attempting to address. Consequently, based on my study of DG-2 materials, I suggest the term 'normal-

956 stress diffusivity'. However, in a pinch one might simply try *mechanical diffusivity* ($\chi = d^2/\tau$), as it

957 provides a descriptive counterpoint to the thermal diffusivity, above. Semantics aside, the crucial thing

958 here is that geologists acknowledge the fundamental theoretical difference between the mode of energy

959 dissipation in rheological fluids, and the possible modes of energy dissipation in rheological solids. The

960 theory discussed here is parameterized by the ratio $\kappa/\chi$, which might be called *thermomechanical*

961 *competence*. Regardless of its name, it provides new insight into the nature of Earth's plate-like blocks,

962 and their relative horizontal and vertical motions.

963

964 7.2 Implications

965 There are good reasons to think that the observed earthquake depth-energy release signal is real. First,

966 while it is known that hypocenter locations are strongly influenced by reference model, centroid locations

967 are not. This is because hypocenter locations are triangulated using body wave travel times, while

968 centroids are located using a summation of long-period body wave and free-oscillation modes. In fact, the

969 relocation algorithm for centroids is remarkably insensitive to small scale *a priori* structure (Aki and

970 Richards, 2002). Furthermore, earthquakes are but one, relatively minor, mode of energy dissipation for

971 the planet, and hence must at some level be related to more general deformation processes in a self-

972 gravitating body (Chao and Gross, 1995; Chao et al., 1995). If the observed coincidence between the

973 earthquake depth-energy release curves and ThERM is real, then it provides crucial support for the

974 hypothesis that Earth's tectonic plates are part of a thermomechanical boundary layer, as much as 700-km

975 thick, that has developed over the course of Earth's history (Patton, 2001; Patton and Watkinson, 2009).

976

977 This coincidence has interesting implications for post-orogenic extensional collapse of orogens, particularly

978 when compared with recent reviews of pressure-temperature-time paths for metamorphic rocks. Maximum

979 pressure estimates, based on equilibrium mineral phase assemblages, for mid-crustal schists and gneisses

980 also commonly coincide with ThERM shear modes (Brown, 2007). A striking example of this is the depth-

981 pressure classification of detachment faults exposed in metamorphic core complexes in the northwestern

982 United States and southwestern Canada, where maximum pressures coincide with H4, L1, and L2 (Patton

983 and Watkinson, in review).

984

985 If observed variations in earthquake depth-energy release are interpreted as a proxies for strength, or more

986 specifically for depth and lateral contrasts in thermomechanical competence ($\kappa/\chi$), then the weakest parts of

987 the crust globally are located between 15- and 25-km depths (Figure 3b). Regionally the weakest crust is

988 found in continental transform, oceanic ridge/transform, continental rift, and Himalayan-type convergent

989 margins (Figures 4, 5). The appearance of ocean crust in this list probably reflects its relative thinness,

990 rather than its intrinsic weakness. The upper and lower boundaries of this weak layer occur at depths

991 consistent with predicted shearing modes H4 and L1 of ThERM (Patton and Watkinson, 2010). Finally,

992 there are additional triplications and slope changes at about 37-km depth, coincident with mode L2 of

993 ThERM, also suggestive of a vertical change in thermomechanical competence. As mentioned above,

994 metamorphic tectonites with pressures of this magnitude are common in collapsed orogens (Brown, 2008).

995 It is likely that the integration of crustal pressure-temperature-time data with centroid moment release

996 studies at a regional scale will be a potentially fruitful avenue for future research.

997

998 Significant differences in the number of hypocenters and centroids occur to about 40-km depth (Table 2).

999 This simply reflects the depth range most affected by the procedural relocation of events in the CMT

1000 catalog. However, tectonically notable differences exist between the map and depth distributions of

1001 hypocenters and centroids above about 25-km depth, roughly coincident with mode L1 of ThERM (Patton

1002 and Watkinson, 2010). Centroids populate both convergent and divergent margins throughout this typical

1003 'crustal' depth range. In contrast, hypocenters largely occur above 15-km depth, with rare events as deep as

1004 25 km at divergent margins. Taken together, these distributions suggest a global maximum seismogenic

1005 crustal thickness of about 25 km (Figure 4a, b).

1006

1007    Using a thermal diffusivity of $10^{-6}$ $m^2s^{-1}$, typical for silicates (Vosteen and Schellschmidt, 2003), the

1008    equivalent thermal age for the 25-km-thick seismogenic crust is about 20 Ma. Presumably, this is the time

1009    needed for decompression melting processes at oceanic ridges to settle down to a steady-state (Crosby et

1010    al., 2006). Recent estimates of pressure and temperature ranges for MORB extraction, based on natural

1011    compositions, are about 1.2-1.5 GPa and 1250-1280 °C (Presnall and Gudfinnsson, 2008). These pressures

1012    correspond to depths of about 40-50 km, consistent with the depth extent of seismic energy release at

1013    oceanic ridges (Figure 4a, orange curve). Furthermore, these workers propose that fracturing of newly

1014    formed lithosphere induces the explosive formation and escape of $CO_2$ vapor, which drives MORB

1015    volcanism, while the source region for material forming the oceanic lithosphere extends no deeper than

1016    about 140 km. In this and other models, the MORB source region therefore lies comfortably below L2.

1017

1018    Hypocenters and centroids are common in the Alpine-Himalayan belt to depths of about 100 km, roughly

1019    coincident with mode L4 of ThERM, with rare events as deep as about 172 km, coincident with mode M1.

1020    Events beneath the Hindu-Kush occur to about 260-km depth, consistent with mode M2. All events deeper

1021    than about 260 km appear to be associated with down going slabs from past or present subduction of

1022    oceanic lithosphere. Note also that anelastic tomography (Romanowicz, 1994) reveals significant lateral

1023    variations in the attenuation of shear waves in the uppermost 250 km of the mantle, which are correlated

1024    with oceanic ridges and continental shields. The shear attenuation pattern below this depth apparently

1025    shifts, correlating with the global distribution of volcanic hotspots. Consequently, regions of Earth's

1026    asthenosphere underlying oceanic ridges and hotspots attenuate more low-frequency seismic energy than

1027    do cratons and shields, which can remain seismically active to similar depths. This is the same depth range

1028    as that potentially impacted by material coupling of surface waves in the thermomechanical boundary layer

1029    of ThERM (Table 5).

1030

1031    Using a typical thermal diffusivity of $10^{-6}$ $m^2s^{-1}$, the equivalent thermal ages for 100-km thick lithosphere,

1032    172-km thick upper tectosphere, and 255-km thick lower tectosphere are about 310 Ma, 940 Ma and 2Ga,

1033    respectively. In contrast, the equivalent *thermal* age for a layer 690-km thick is about 15 Ga, more than

1034    three times the age of the Earth and comparable to the age of the universe. Clearly, it is implausible that 1035    the structure of the asthensphere is solely due to thermal diffusion processes.  Consequently, some dynamic

1036    convective (advective?) process is needed to maintain differential stresses to these depths (McKenzie,

1037    1967).

1038

1039    Geologic history has been divided into ocean basin time (0-200 Ma), plate tectonic time (200-950 Ma), and

1040    "pre-tectonic" time (950-2300 Ma), on the basis of marine and continental geology (Moores and Twiss,

1041    1995).  Also, based on Brown's (2008) recent review, the equivalent thermal ages, above, correspond to the

1042    amalgamation times for supercontinents Pangea, Rodinia, and Nuna.  Reconciling the global smoothness of

1043    thermal (and viscous) diffusion with the spatial complexity and local non-smoothness of tectonic

1044    deformation processes requires a model that incorporates a self-consistent mechanism for strain localization

1045    with depth.  Isobaric shearing in the asthenosphere offers a more comprehensive explanation for these first-

1046    order observations, than does the standard Earth model.  It is time for the strength of solid earth materials to

1047    be included in dynamic plate theory.

1048

1049    As noted in Section 1.0, the spectral characteristics of Earth's gravity-topography correlation and

1050    admittance, combined with the adiabaticity of the lower mantle, are consistent with vigorous convection in

1051    the sublithospheric mantle of the standard Earth model, at lateral scales $\lambda_{\frac{1}{2}} > 1000\ km$.  This conclusion

1052    complements those of regional isostasy, where near surface loads are supported by the flexural rigidity of

1053    the crust and lithosphere, at lateral scales $\lambda_{\frac{1}{2}} < 1000\ km$.  This latter range of wavelengths, down to about

1054    50 km, has been called the 'diagnostic waveband of flexure' ((Watts, 2001), pg. 178).  Furthermore, the

1055    inviscid (i.e., mechanically indeterminate) nature of the asthenosphere in flexural isostasy does not conflict

1056    with the fluid nature of the sublithospheric mantle in the standard model.  Consequently, it is reasonable to

1057    conclude that mantle convection actively supports Earth's longest-wavelength gravity and geoid anomalies

1058    (Hager, 1984; McKenzie, 1967; Panasyuk and Hager, 2000; Richards and Hager, 1984; Steinberger et al.,

1059    2010).  Based on recent crustal thickness and flexural modeling, Steinberger *et al* (2010) suggest that

1060    Earth's gravity anomalies with $\lambda_{\frac{1}{2}} \geq 650\ km$ ($l \leq 30$) are probably due to sources in the sublithospheric

1061    mantle, while those at shorter wavelengths have sources predominantly in the lithosphere.  This downward

1062    revision of the upper limit for the diagnostic waveband therefore provides more room for the operation of

1063    the standard model, including mantle plumes.

1064

1065    Given the presence of a thermomechanical boundary layer at Earth's surface, some 700-km thick, as

1066    suggested by this study, the interpretation of gravity-topography spectra should be revisited.  Based on

1067    earlier incipient modes analysis, a competent layer near the surface of a dynamic thermomechanical planet

1068    can be expected to develop low-amplitude material waves with wavelengths ranging from 2 to $\zeta$ times the

1069    layer thickness.  Assuming the depths of the isobaric shears (Table 3) define a series of layer thicknesses,

1070    the equivalent ranges of angular degree can be worked out (Table 5).  Coincidentally, the long-wavelength

1071    limit for folding of a layer M2 thick is $l = 22$, roughly coincident with the roll off in the gravity-topography

1072    correlation (Wieczorek, 2007).  Furthermore, the empirical limit suggested by Steinberger *et al* (2010) is

1073    bracketed by the short-wavelength M4 and the long-wavelength M1 limits, and coincident with the short-

1074    wavelength M3 limit.  In other words, the correlation they propose is supported by these findings, provided

1075    that the lithosphere is taken to include the continental tectosphere.  However, the implications of

1076    thermomechanical theory for convection in the deep mantle are dramatically different from those of the

1077    standard model.  For example, the long-wavelength M3 and M4 limits are $l = 8$ (Table 5).  Consequently,

1078    gravity and geoid anomalies only with $\lambda_{\frac{1}{2}} > 2500\ km$ ($l < 8$) can be unequivocally associated with lower

1079    mantle convection.  Coincidentally, this is the scale of robust lateral variations imaged by seismic

1080    tomography (Dziewonski et al., 1977; Dziewonski and Woodward, 1992; Gu et al., 2001).  Furthermore,

1081    given the inverse temperature dependence of thermal expansivity for inhomogeneous elastic solids, any

1082    convection process in the lower mantle is likely to be rather sluggish.  Detailed study of these intriguing

1083    spectral observations is in progress.  Finally, material coupling in these layers is likely to have a measurable

1084    impact on the attenuation of intermediate and long-period surface waves, consistent with the predicted non-

1085    adiabaticity of the upper mantle (Durek and Ekstrom, 1996; Romanowicz, 1995).

1086

1087    **8.0 Conclusion**

1088    Global plate motions are the result of a complex planetary-scale rock mechanics experiment, where the

1089    motive force of gravity drives thermomechanically competent oceanic lithosphere into the relatively

1090    weaker rock units of the continents. Under extreme pressures, isobaric shearing at discrete depths

1091    facilitates toroidal plate motions, i.e., Euler rotations of spheroidal caps, while minimizing global energy

1092    dissipation. The fractal depth distribution of these disclinations is dictated by the statistically-stable

1093    thermomechanics of shear localization in inhomogeneous non-linear elastic self-gravitating solids, and

1094    scales with the thickness of mature oceanic lithosphere, where the greatest density contrasts and

1095    gravitational body forces reside. At low pressures, shear localization in crustal rocks occurs as dislocations

1096    at finite angles with respect to the shortening direction, with a 30 degree angle being the most likely.

1097    Consequently, relatively low-angle (~30º) reverse faults, steep (~60º) normal faults, and triple junctions

1098    with orthogonal or hexagonal symmetries are likely to form in regions of crustal shortening, extension, and

1099    transverse motion, respectively. In convergent plate boundary regions, this results in the overall thickening

1100    and seismogenic deformation of weaker rocks in the zone. Once convergence ceases, these regional high-

1101    potential energy welts relax over 30 Ma periods, often exposing equilibrium assemblages of middle- and

1102    lower-crustal rocks at the surface which are bounded by crustal scale shear zones. Deep crustal migmatites

1103    are produced by dehydration anatexis during shortening, and heat release following gravitational collapse

1104    drives further melting and differentiation of the crust. Equilibrium pressures in these crystalline cores are

1105    consistent with exhumation from 25- and 37-km depths, coincident with the $L1$ and $L2$ isobaric shears of

1106    ThERM. At extensional plate margins, brittle failure of crustal rocks depressurizes the subjacent mantle,

1107    resulting in its partial fusion and the subsequent eruption of basaltic lavas. These lavas cool and solidify,

1108    preserving a record of the ambient magnetic field. At transform margins, seismic energy dissipation is

1109    limited to about 25-km depth, but can be distributed over a wide region, depending on the competence of

1110    the rocks on either side of the fault. Wider shear zones are likely in continental regions, because of the

1111    lower competence of felsic crust. The dominance of toroidal motions at Earth's surface can be attributed to

1112    the combination of liquid water, which has a profound weakening effect on common rocks and minerals,

1113    and the localization of shear at discrete depths within the thermomechanical Earth. The absence of plate

1114    tectonics on other terrestrial planets is simply due to the small number of known examples. Furthermore,

1115    the correlation and admittance of Earth's gravity-topography spectra can be reconciled with this novel

1116    thermomechanical theory. In consequence, mantle convection, in the sense of the standard model, is

1117    possible only in the deep mantle at a scale $\lambda_{\frac{1}{2}} > 2500\ km$ ($l < 8$), consistent with the scale of robust lateral 1118    variations imaged by seismic tomography.  Finally, the temperature dependence of thermal expansivity for

1119    this thermomechanical Earth makes vigorous convection in the lower mantle unlikely.  The dominant mode

1120    of thermal convection for the planet, therefore, looks and acts like a dynamic version of plate tectonics.

1121

1122    **Acknowledgements.**  The figures were prepared using the Generic Mapping Tools (Wessel and Smith,

1123    1998).  I thank Professor A. John Watkinson for his unflagging support and quiet insistence that structural

1124    observations at all scales matter, Professor Bruce E. Hobbs for his support and leadership on matters

1125    thermodynamical, Professor Raymond Joesten for pointing out the coincidence of planar tectonite fabrics in

1126    mantle xenoliths from depths consistent with mode M1, and Professor Frank Horowitz for suggesting the

1127    least-squares fitting of ThERM to PREM.  This paper is dedicated to the memory of Dr. Peter Hornby.

*Figure captions*

**Figure 1.** Plots of a) hypocenter and b) centroid depths for the 20646 algorithmically relocated earthquakes from the global CMT catalog (Ekstrom and Nettles, 2011), with origin times in the period January 1976 through December 2010. With increasing depth, the color breaks coincide with the isobaric shears H4, L1, L2, L4, M1, M2, and M4, of the ThERM (Patton and Watkinson, 2009, 2010). The CMT relocation procedure generally maps crustal seismicity to deeper levels, as indicated by the contrasting orange and yellow hues of ocean ridge earthquakes.

**Figure 2.** Plots of centroid depths for earthquakes in six tectonic settings (Table 4):
a) continental transform; b) Himalaya-type convergence; c) oceanic ridge/transform; d) island arc-type convergence; e) continental rift; and f) Andean-type convergence.
These subsets of the CMT catalog are used to compute the earthquake depth-energy release curves appearing in Figures 3-5.

**Figure 3.** Plots of seismic depth-energy release $\Sigma M_W(z;3)$, smoothed using a 3-km thick boxcar filter and color-keyed by tectonic setting (Table 4, Figure 2): a) Hypocenter depth-release curves exhibit sharp peaks at about 10- and 30-km depths and appear somewhat artificial, most likely due to the quick-epicenter location routines used by the USGS and ISC; b) Centroid depth-release curves span a broad range of depths and exhibit peak energy release at 17- to 18-km depths. The peak amplitude of most curves is proportional to the number of earthquakes in that tectonic setting. About ¾ of the earthquakes in this catalog occur in island arc-type convergence zones.

**Figure 4.** Plots of seismic depth-energy release $\Sigma M_W(z;3)/N$, smoothed using a 3-km thick boxcar filter, normalized by the number of events $N$, and color-keyed by tectonic setting (Table 4, Figure 2): a) depth-release curves at divergent and transform margins extend no deeper than about 50 km, and exhibit large positive and negative excursions from the global average at typical crustal levels. The deepest earthquakes coincide with the MORB source region (Presnall and Gudfinnsson, 2008); b) depth-release

curves at convergent margins extend to depths of about 700 km, and also exhibit large positive and negative excursions from the global average at typical crustal levels.

**Figure 5.** Color-keyed plots of seismic energy release with depth $\Sigma M_w\left(z;10\right)/N$, normalized by the number of events $N$ in convergent margins, and smoothed with a boxcar filter 10-km thick. While the seismic energy release with depth in island arc-type convergent margins (purple) closely approximates the global baseline (black dashed), the Himalaya-type (green) and Andean-type (blue) margins display marked deviations in depth ranges. Positive and negative excursions of these curves likely indicate lateral variations mantle strength, and suggest stratification consistent with modes L4, M1, and M2 of ThERM (Patton and Watkinson, 2009).

**Figure 6.** Graphical summary of the statistical thermodynamics of strained inhomogeneous elastic and self-gravitating matter configurations: a) mechanical variability bears little insight for DG-2 materials; b) statistical variability offers crucial insight for the thermodynamics of shear localization in DG-2 materials

**Figure 7.** Correlation of experimental data from fourteen samples of the Mt. Scott granite, subjected to load-hold analysis (Katz and Reches, 2002, 2004), with the predictions of statistical thermodynamics and deformation modes of DG-2 materials: a) Normalized differential stresses (NDS) on intact samples ($\psi = \zeta$ *NDS*) are plotted as functions of diameter/length (*d/l*) for load-hold (green circles, red diamonds) and load-to-failure tests (orange diamonds). Three samples (red diamonds) spontaneously failed during the designated hold period. All macroscopically failed samples (diamonds) plot above the localization threshold curve ($\psi^L$, orange) given the range of standard error in Coulomb strength for the Mt. Scott granite ($586 \pm 16$ MPa, dashed curves, inset); b) Populations of microscopic cracks with respect to the loading direction, observed post-loading, fall in the range $\frac{1}{2} < \kappa/\chi < 2.9$, while macroscopic shears fall in the range $0.6 < \kappa/\chi < 0.9$. The thermomechanically rigid portion of this system is the stiff load frame itself, which by definition plots at $\kappa/\chi = 1$. The wide range of microscopic crack angles probably reflects the prior geometry of sample grain-size and fabric.

**Figure 8.** Comparison of ThERM (Patton and Watkinson, 2009), scaled to a fundamental thickness of 99.54 km representing mature oceanic lithosphere, with depth variations in compressional $V_p$ and shear $V_s$ wave speeds, density $\rho$ (Dziewonski and Anderson, 1981), and normalized seismic depth-energy release $\Sigma M_w(z)/N$ (Figure 4b). The thickness of cratons and shields (Artemieva and Mooney, 2001) correlate with the M1 and M2 shears of ThERM (Patton and Watkinson, 2009), while the depth cut-off of seismicity, at about 700-km depth (Frohlich, 1989), correlates with M3-4. Given that thermomechanical competence generally decreases with depth in ThERM, subducting slabs are likely to freely enter the lower mantle, in contrast to the mesosphere hypothesis (Isacks and Molnar, 1969; Isacks et al., 1968). The re-orientation of pressure and tension axes in deep earthquake focal mechanisms are therefore likely to result from anti-crack shear ruptures in metastable spinel mineral species at very high pressures (Green and Burnley, 1989). The depth cut-off of seismicity might result from a rapid decrease in entropy density at 700-km depth (Patton and Watkinson, in review).

**Figure 9.** Comparison of the crustal 'overtones' of ThERM (Patton and Watkinson, 2009), scaled to a thickness of $F\zeta^{-1} = 14.4\ km$ representing the brittle crust, with depth variations in compressional $V_p$ and shear $V_s$ wave speeds, density $\rho$ (Dziewonski and Anderson, 1981), and normalized seismic depth-energy release $\Sigma M_w(z)/N$ (Figure 4). The greatest variations of earthquake depth-energy release correlate with the low-velocity zone, as well as several distinct levels in the crust, lithosphere, and tectosphere. In regions where the tectosphere is absent, the asthenosphere resides subjacent to the crust.

**a)**

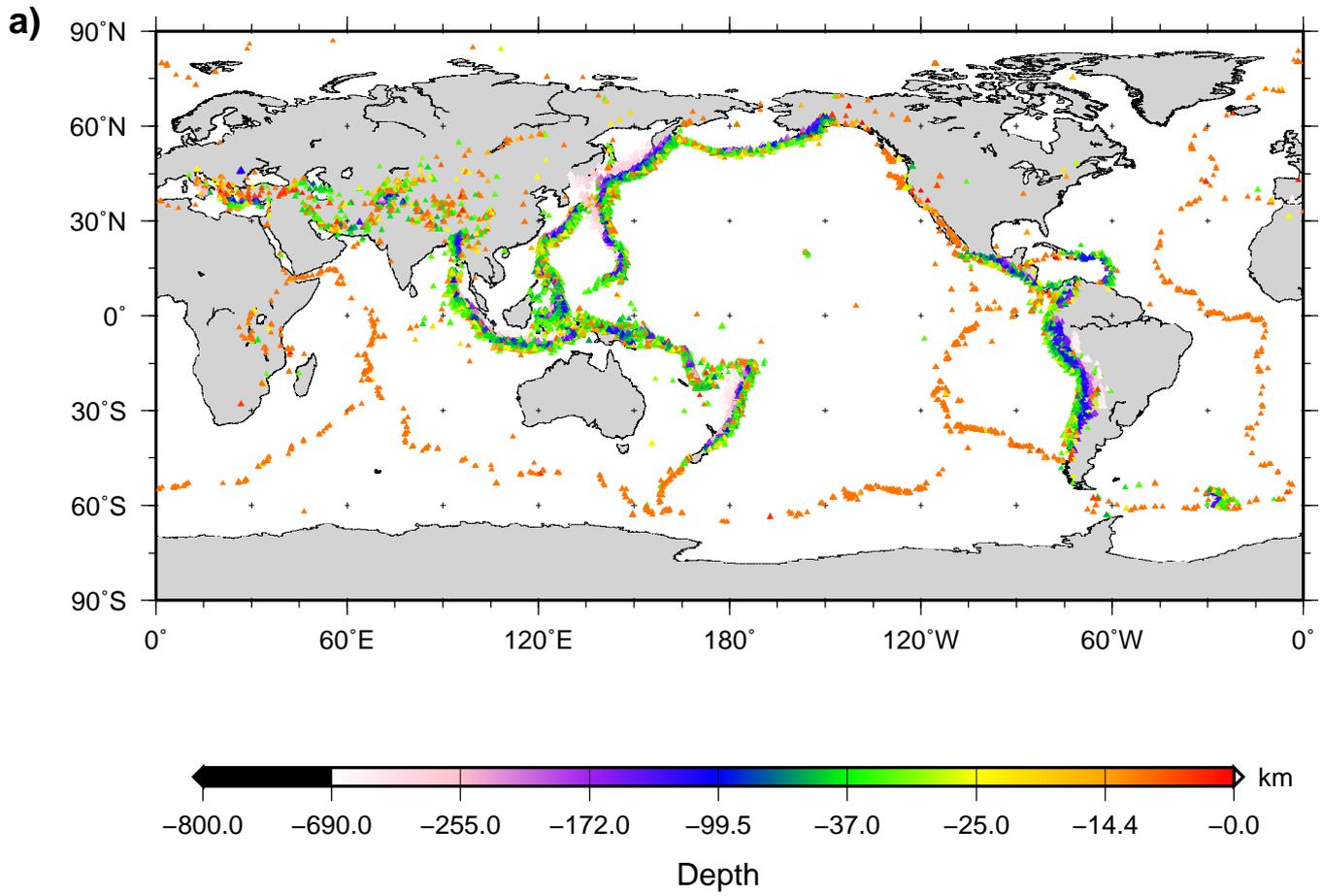

**b)**

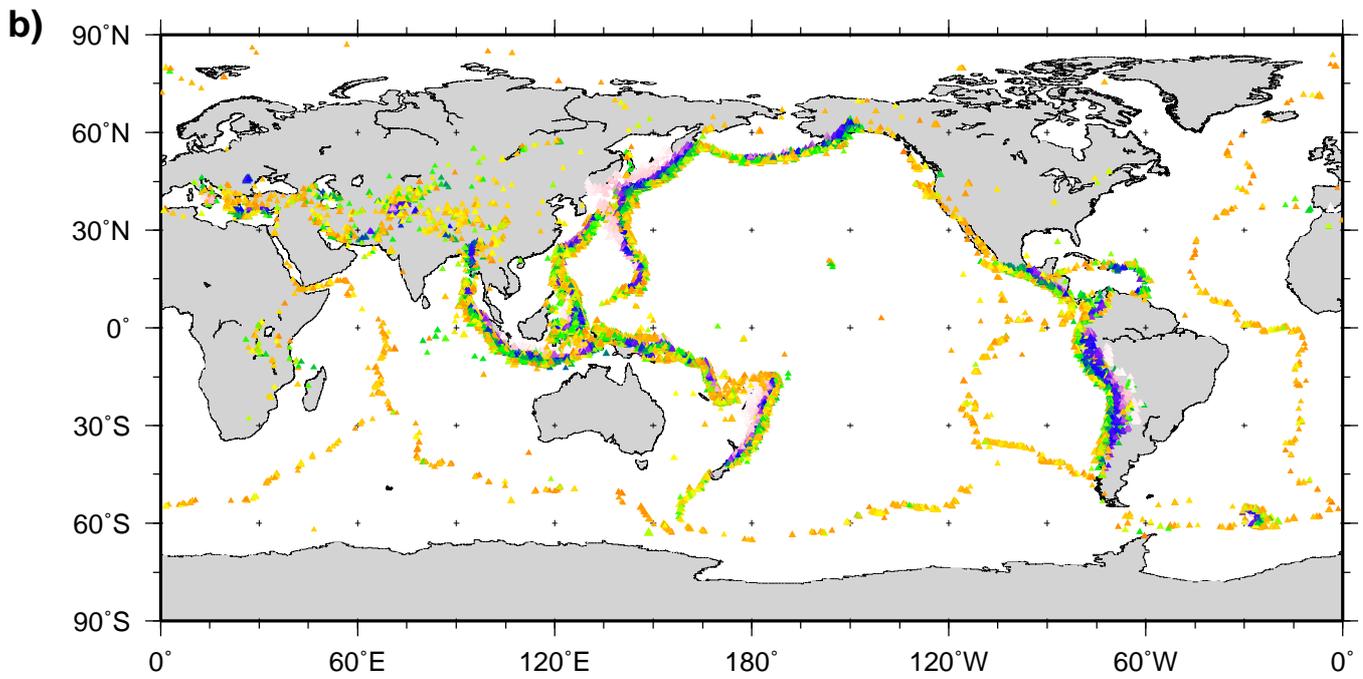

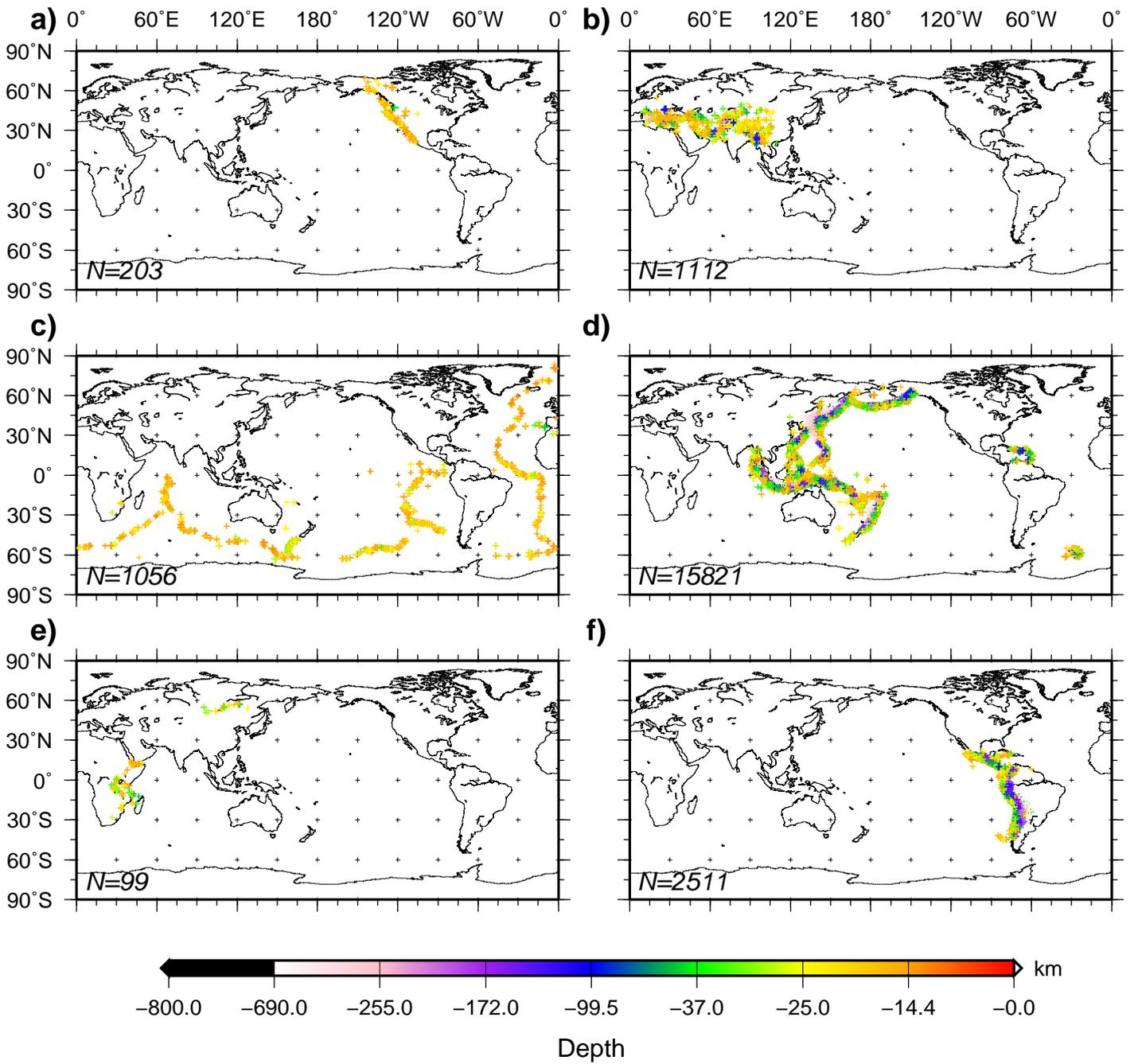

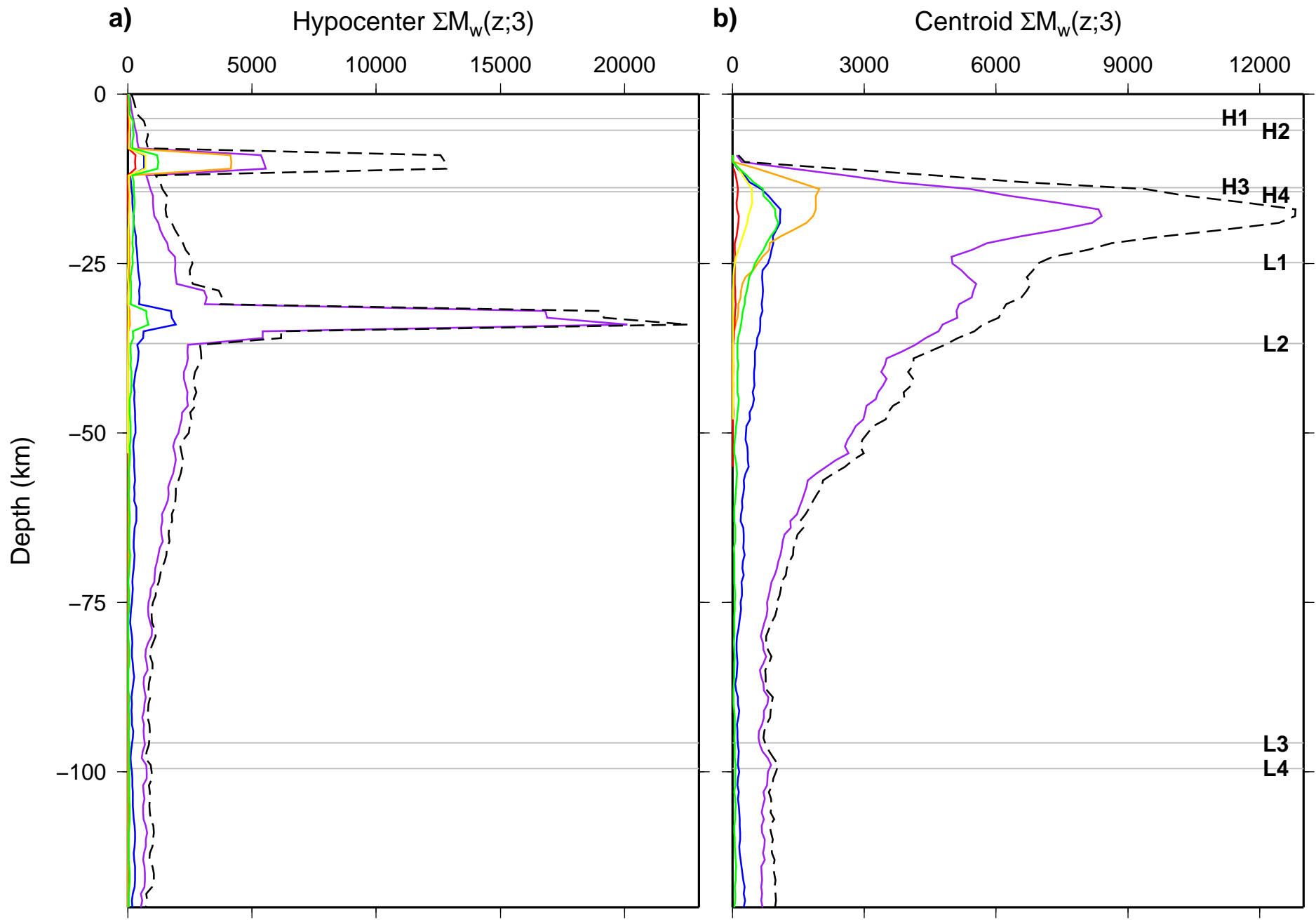

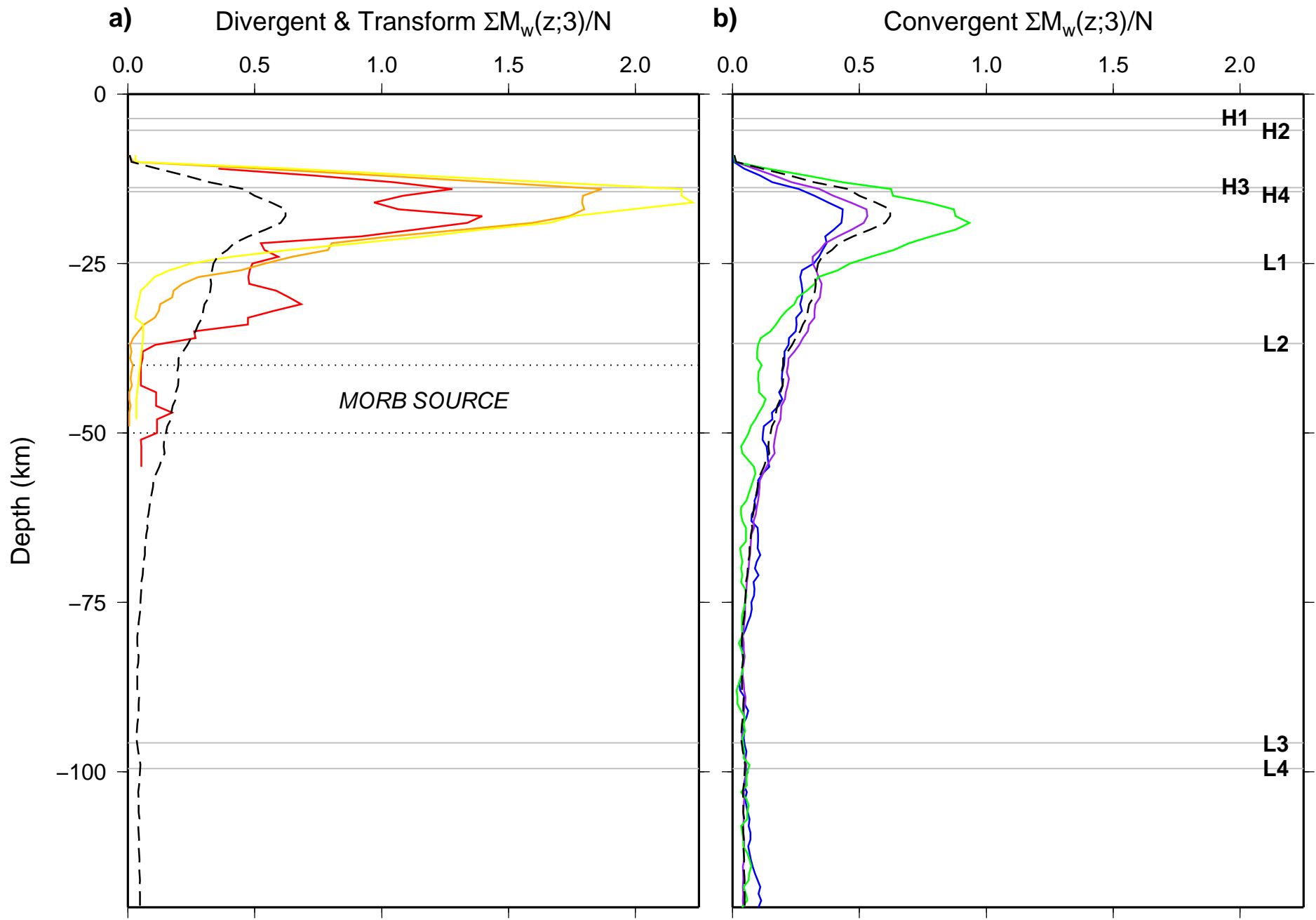

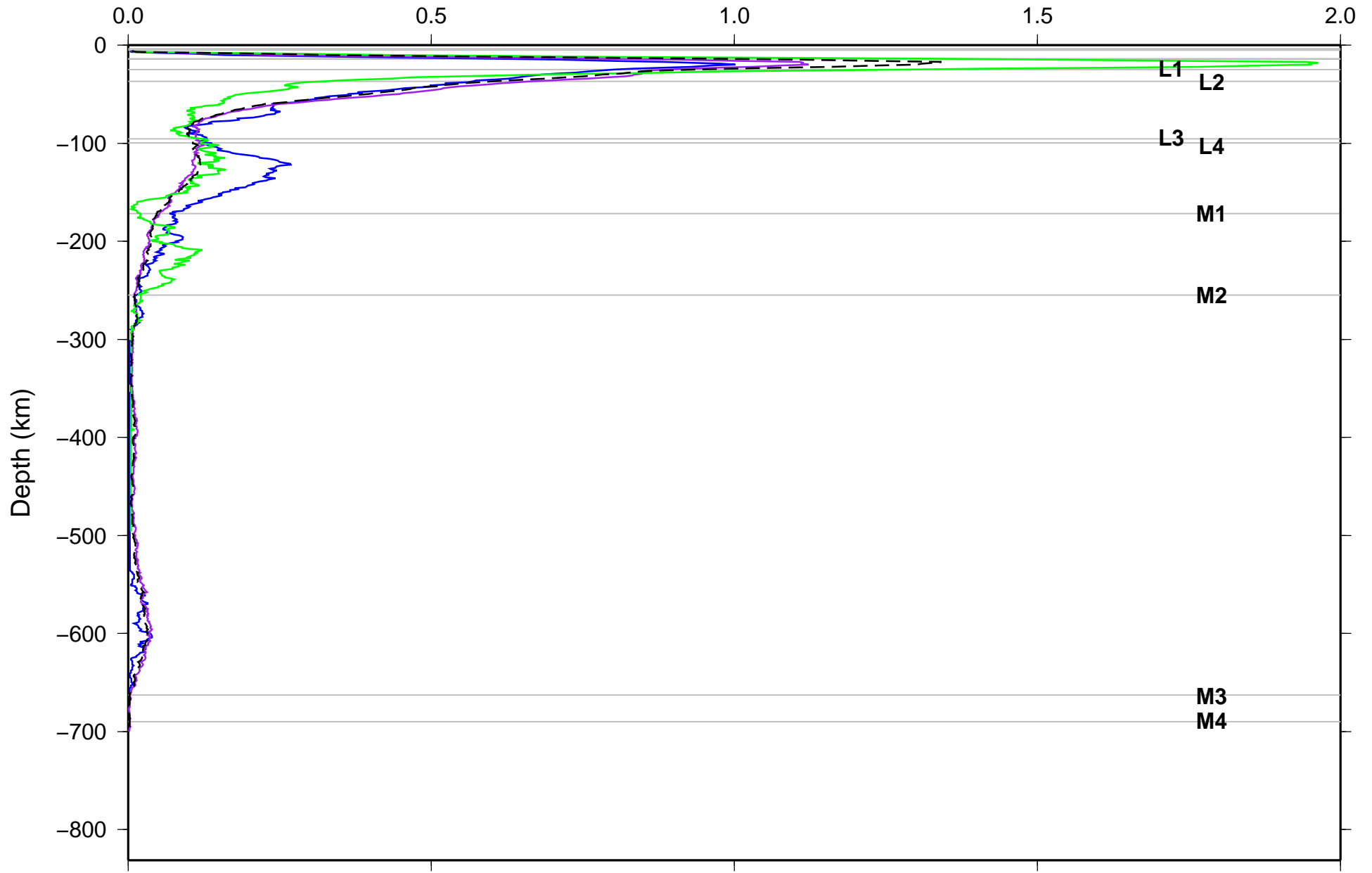

**a)**

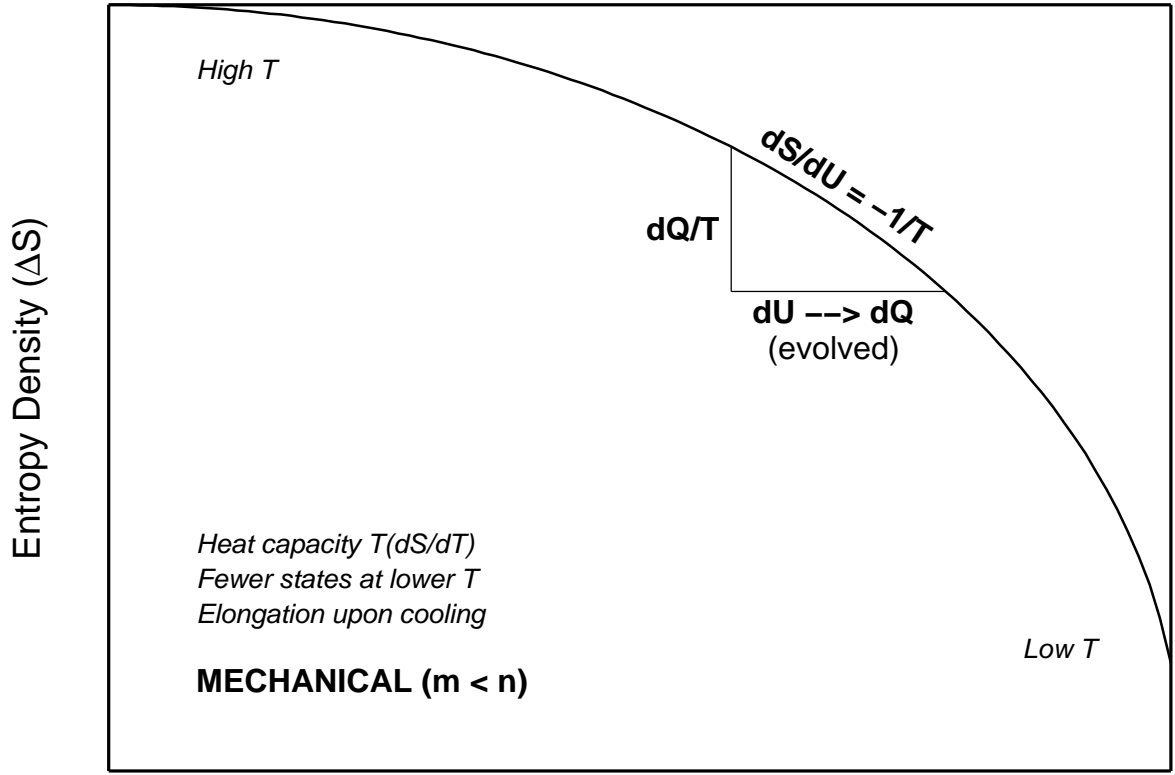

**b)**

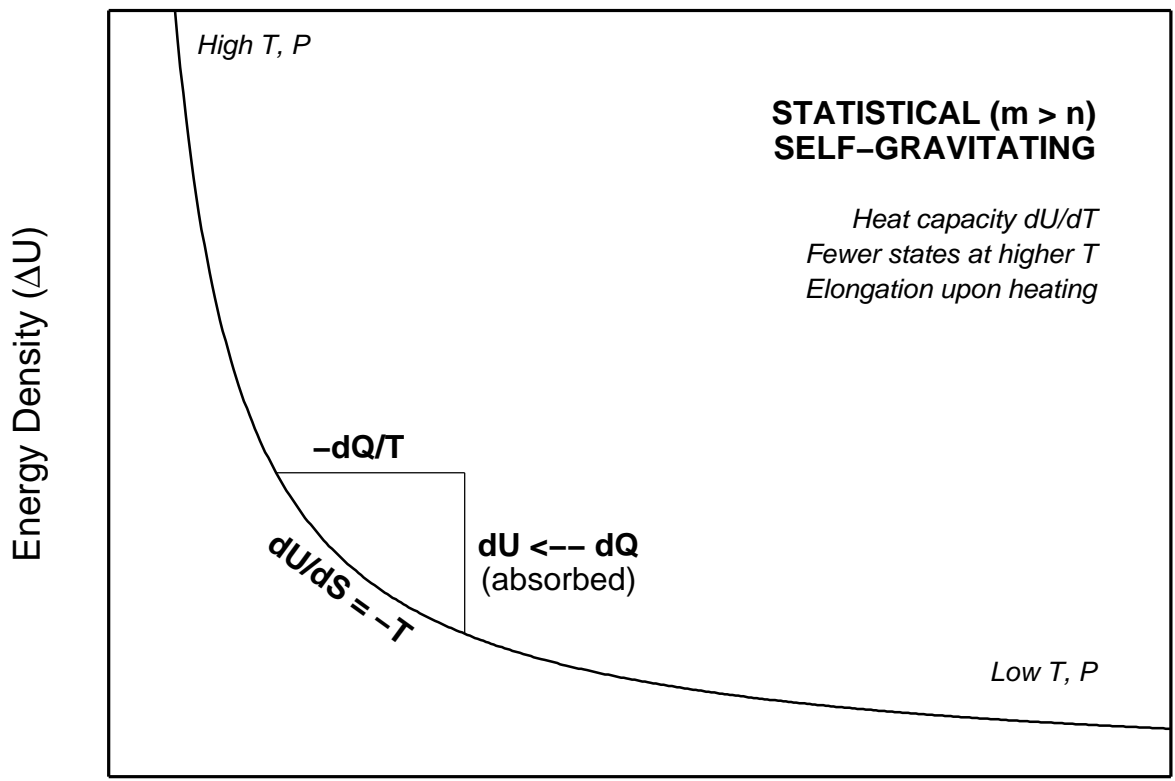

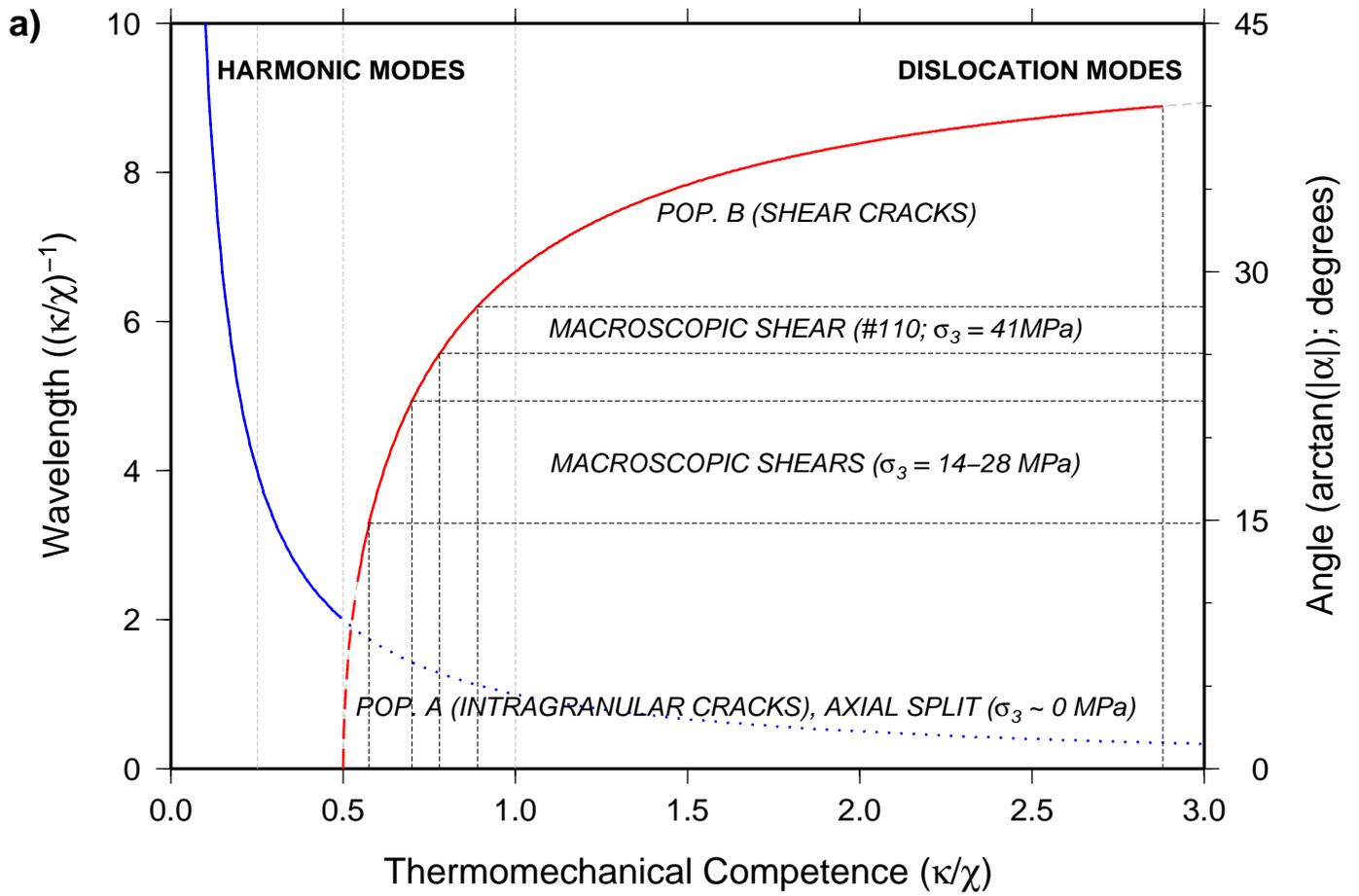

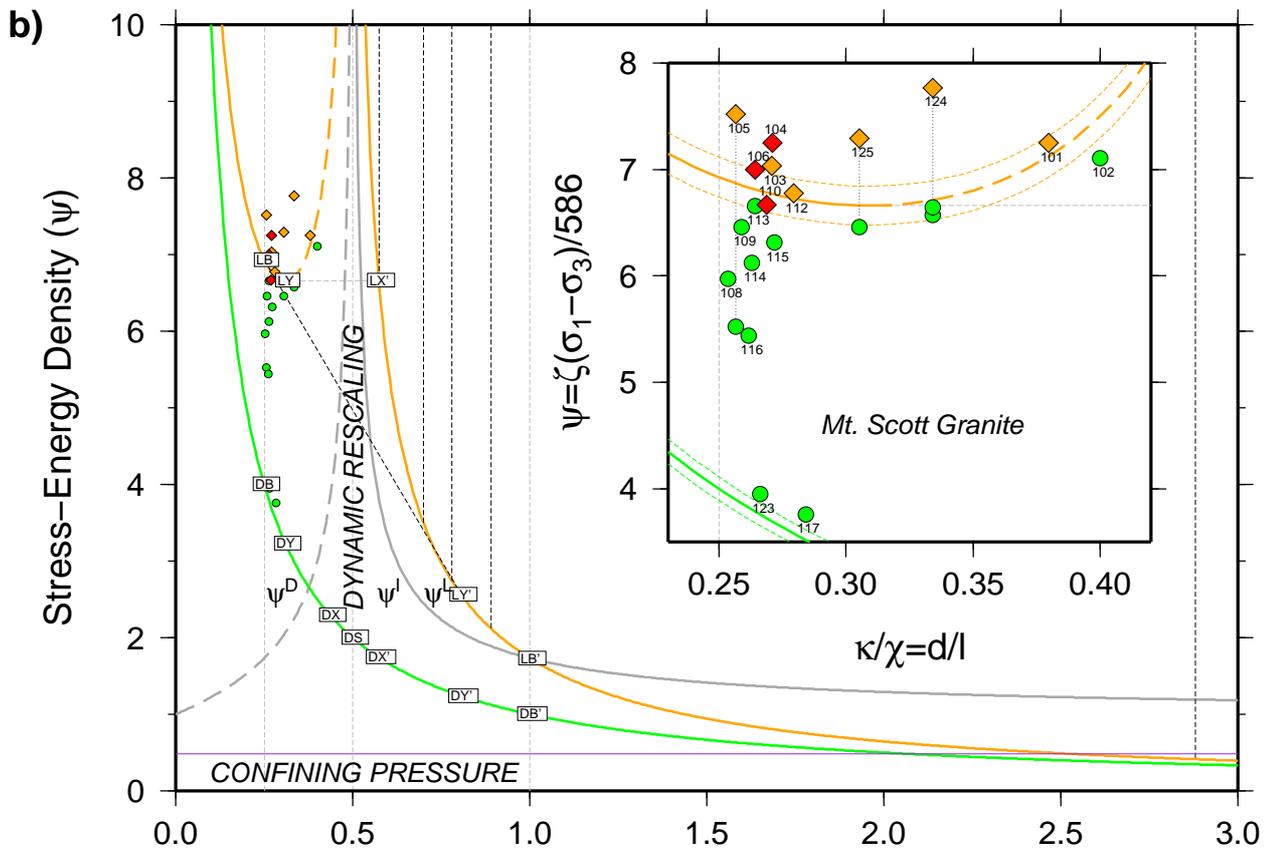

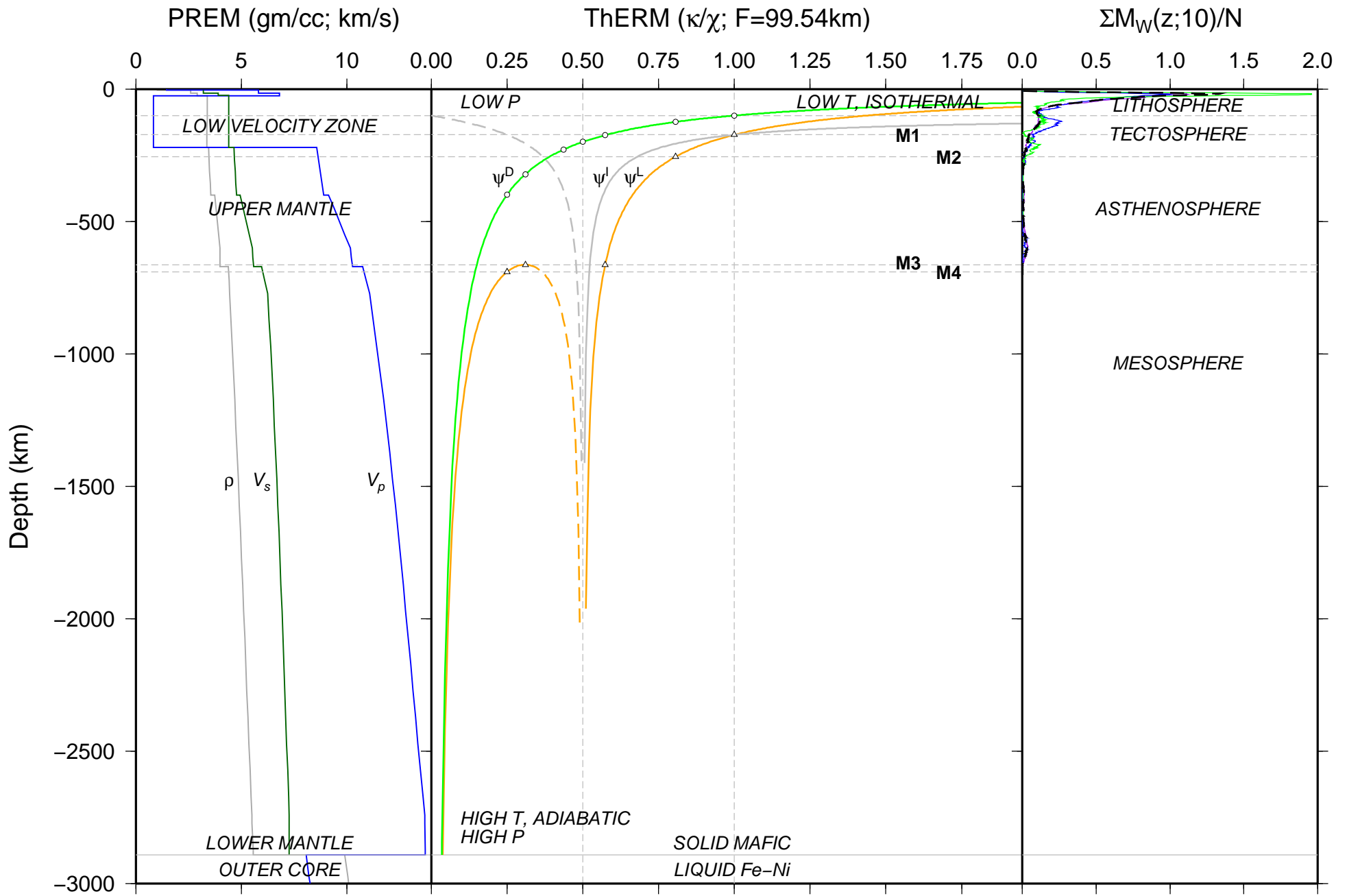

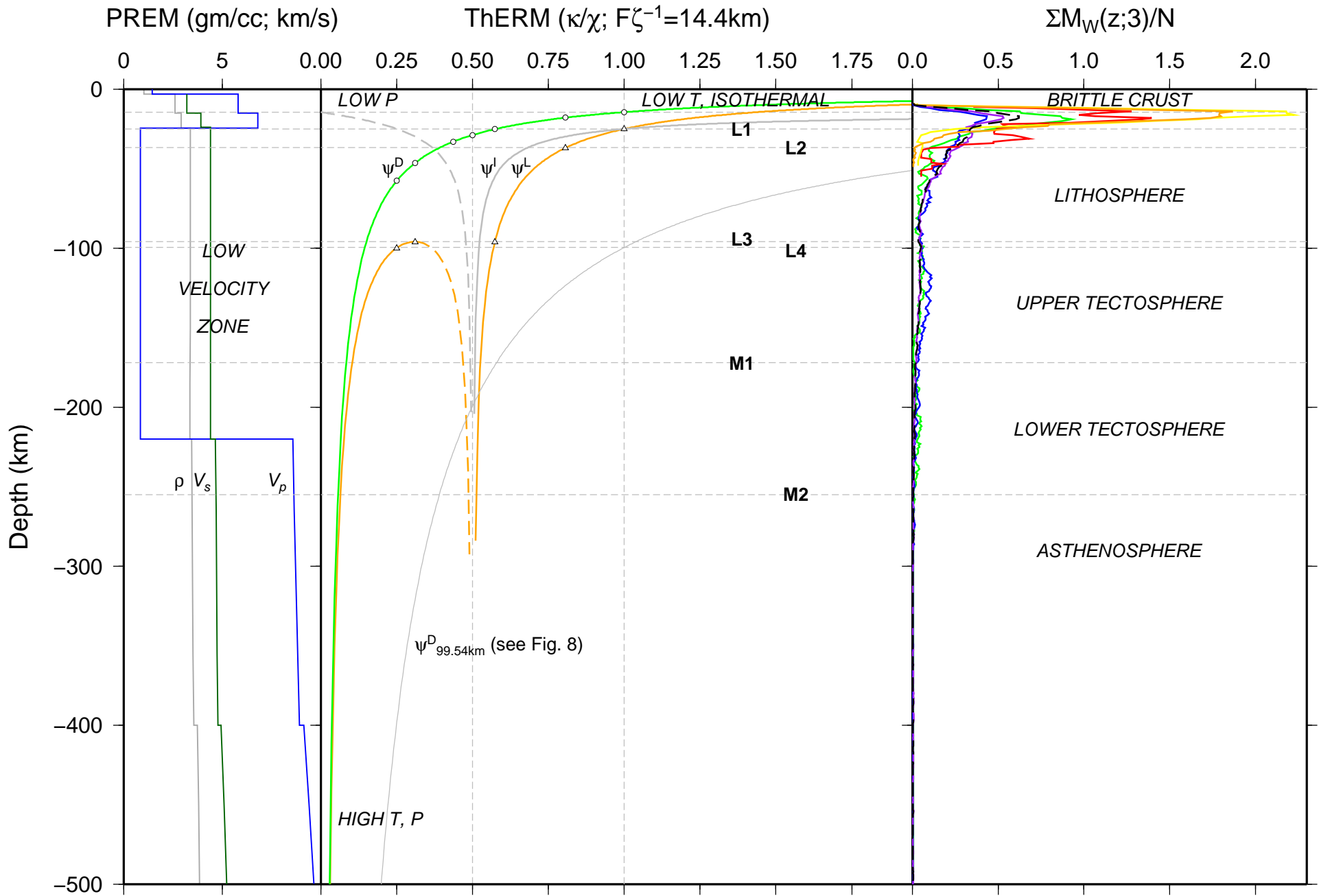

**Table 1.** Hypocenter and centroid statistics by magnitude

| $M_w$ range | Hypocenters | Centroids | Difference |
|---|---|---|---|
| 0-1 | 199 | - | (199) |
| 1-2 | - | - | - |
| 2-3 | - | - | - |
| 3-4 | 4 | - | (4) |
| 4-5 | 3849 | 2102 | (1747) |
| 5-6 | 15262 | 16053 | 791 |
| 6-7 | 1317 | 2182 | 865 |
| 7-8 | 14 | 292 | 278 |
| 8-9 | 1 | 16 | 15 |
| 9-10 | - | 1 | 1 |
| Total | 20646 | 20646 | 0 |

**Table 2.** Hypocenter and centroid statistics by depth

| Depth range | Hypocenters | Centroids | Difference | Color range |
|---|---|---|---|---|
| 0-14.4 | 3127 | 1031 | (2096) | red-orange |
| 14.4-25 | 1051 | 5268 | 4217 | orange-yellow |
| 25-37 | 5667 | 3471 | (2196) | yellow-green |
| 37-99.5 | 5511 | 5495 | (16) | green-blue |
| 99.5-172 | 2520 | 2605 | 85 | blue-purple |
| 172-255 | 941 | 929 | (12) | purple-pink |
| 255-690 | 1828 | 1838 | 10 | pink-white |
| 690-800 | 1 | 9 | 8 | black |
| Total | 20646 | 20646 | 0 | |

**Table 3.** Depths to isobaric shears of ThERM (Patton and Watkinson, 2010)

| Shear | Depth[*] (km) | Pressure[**] (GPa) |
|-------|---------------|---------------------|
| H1 | 3.6 | 0.10 |
| H2 | 5.3 | 0.14 |
| H3 | 13.8 | 0.37 |
| H4 | 14.4 | 0.39 |
| L1 | 25 | 0.67 |
| L2 | 37 | 1.04 |
| L3 | 96 | 2.96 |
| L4 | 100 | 3.00 |
| M1 | 172 | 5.40 |
| M2 | 255 | 8.27 |
| M3 | 663 | 23.4 |
| M4 | 690 | 24.5 |

[*]$F = 99.54$ km; $\zeta = 4\sqrt{3}$

[**]Based on least-squares misfit to PREM (Dziewonski and Anderson, 1981)

**Table 4.** Centroid-moment magnitude statistics[*] by tectonic setting

| Setting | N | $\Sigma M_W$ | Max. Z (km) | Plot color |
|---|---|---|---|---|
| All relocated events | (20646) | (112877) | 699 | black-dashed |
| Continental transform | 203 | 1094 | 47 | yellow |
| Oceanic ridge/transform | 1056 | 5610 | 49 | orange |
| Continental rift | 99 | 2291 | 53 | red |
| Himalaya-type convergence | 1112 | 5957 | 492 | green |
| Island arc-type convergence | 15821 | 86879 | 699 | purple |
| Andean-type convergence | 2511 | 13887 | 656 | blue |
| Total (0.8% over sample) | 156 | 2841 | | |

[*]Correlation coefficient for (N, $\Sigma M_W$) population is r = 0.999899, with intercept y = -64.4

**Table 5.** Predicted ranges of spherical harmonic degree and frequency for material waves propagating in the boundary layer structure of a thermomechanical Earth

| Layer | Depth (km)[*] | Degree[**] ($l$) | | Spheroidal[†] (mHz) | | Toroidal[‡] (mHz) | |
|---|---|---|---|---|---|---|---|
| | $d$ | $\lambda = \zeta d$ | $\lambda = 2d$ | Low | High | Low | High |
| L1 | 25 | 231 | 794 | 25.6 | - | 23.0 | - |
| L2 | 37 | 156 | 538 | 15.5 | - | 15.3 | - |
| L3 | 96 | 60 | 208 | 6.50 | 21.0 | 7.13 | 23 |
| L4 | 100 | 57 | 200 | 6.28 | 20.0 | 6.93 | 22.4 |
| M1 | 172 | 34 | 116 | 4.24 | 12.0 | 4.38 | 13.2 |
| M2 | 255 | 22 | 79 | 3.08 | 8.43 | 3.00 | 9.20 |
| M3 | 663 | 8 | 29 | 1.43 | 3.75 | 1.35 | 3.74 |
| M4 | 690 | 8 | 30 | 1.43 | 3.85 | 1.35 | 4.00 |

[*]Isobaric shears of ThERM (Patton and Watkinson, 2009, 2010) with F=99.54 km

[**]From Jean's relation $l = \left(2\pi R_{\oplus} / \lambda\right) - \left(1/2\right); R_{\oplus} = 6371 km$

[†]Graphical estimates from Figure 8.8 of (Aki and Richards, 2002)

[‡]Graphical estimates from Figure 8.7 of (Aki and Richards, 2002)

**Table 6.**  Experimental results of load-hold tests on Mt. Scott granite (from (Katz and Reches, 2002, 2004)), with estimates of mechanical diffusivity.  Sample diameter $d = 25.4\ mm$.

| Sample | $l$ (mm) | $\tau$ (min) | Max. NDS | -log $\chi^{*}$ | Comments |
|--------|----------|--------------|----------|-----------------|----------|
| 101 | 66.9 | - | 1.05 | - | load to failure |
| 102 | 63.5 | 95 | 1.03 | - | load hold |
| 103 | 93.8 | - | 1.02 | - | load to failure |
| 104 | 93.7 | 61 | 1.05 | 5.6 | *spontaneous failure* |
| 105 | 99.0 | 180 | 0.80 | - | load hold (1) |
| → | → | - | 1.09 | - | load to failure |
| 106 | 96.1 | 1.25 | 1.01 | 3.9 | *spontaneous failure* |
| 108 | 100.2 | 180 | 0.86 | - | load hold |
| 109 | 98.1 | 180 | 0.93 | - | load hold |
| 110 | 94.5 | 0.03 | 0.96 | 2.3 | *spontaneous failure* |
| 112 | 90.9 | - | 0.98 | - | load to failure |
| 113 | 96.1 | 180 | 0.96 | - | load hold |
| 114 | 96.6 | 180 | 0.88 | - | load hold |
| 115 | 93.4 | 360 | 0.91 | - | load hold |
| 116 | 97.1 | 180 | 0.78 | - | load hold |
| 117 | 89.4 | 180 | 0.54 | - | load hold |
| 123 | 95.4 | 180 | 0.57 | - | load hold |
| 124 | 76.0 | - | 0.95 | - | load hold (1) |
| → | → | - | 0.96 | - | load hold (2) |
| → | → | - | 1.12 | - | load to failure |
| 125 | 83.2 | 180 | 0.93 | - | load hold (1) |
| → | → | - | 1.05 | - | load to failure |

$^{*}$Computed using $\chi = l^{2}/\tau$